\documentclass[11pt,letterpaper]{article}
\ifdefined\StandardPreambleLoaded\expandafter \fi
\newcommand{\StandardPreambleLoaded}{1.0.1}

\usepackage[T1]{fontenc}
\usepackage[utf8]{inputenc}
\usepackage{lmodern}
\usepackage{microtype}
\usepackage[margin=1in]{geometry}
\usepackage{slantsc}
\AtBeginDocument{%
  \DeclareFontShape{T1}{lmr}{m}{scit}{<->ssub*lmr/m/scsl}{}%
}

\usepackage{mathtools,amssymb,amsthm,amscd,aliascnt}
\usepackage[mathscr]{euscript}
\usepackage{bbm,dsfont}
\usepackage{scalefnt}
\numberwithin{equation}{section}

\usepackage[dvipsnames]{xcolor}
\usepackage{graphicx,tikz,pdfpages}
\usepackage{algorithm,algpseudocode}

\usepackage[inline]{enumitem}

\newlist{casesenum}{enumerate}{2}
\setlist[casesenum,1]{label=\textbf{Case~\arabic*.},itemindent=*,leftmargin=0pt}
\setlist[casesenum,2]{label=\textbf{Case~\roman*.},itemindent=*,leftmargin=\parindent}
\usepackage[
  backend=biber,
  style=alphabetic,
  sorting=nyt,
  maxnames=8,
  maxalphanames=8,
  backref=true,
  url=false
]{biblatex}
\usepackage{subfiles}

\IfFileExists{support/commenting.sty}{\usepackage[draft]{support/commenting}}{}

\definecolor{DocumentLink}{RGB}{25,62,105}
\usepackage[colorlinks=true,linkcolor=DocumentLink,citecolor=DocumentLink,urlcolor=DocumentLink]{hyperref}
\usepackage[nameinlink,noabbrev,capitalize]{cleveref}
\crefname{paragraph}{Paragraph}{Paragraphs}

\newcounter{algorithmicblock}
\AddToHook{env/algorithmic/begin}{\stepcounter{algorithmicblock}}
\makeatletter
\providecommand{\theHALG@line}{}
\renewcommand{\theHALG@line}{\arabic{algorithmicblock}.\arabic{ALG@line}}
\makeatother

\ifdefined\StandardTheoremsLoaded\expandafter \fi
\newcommand{\StandardTheoremsLoaded}{1.0.0}

\newcommand{\DeclareSharedTheorem}[2]{%
  \newaliascnt{#1}{theorem}%
  \newtheorem{#1}[#1]{#2}%
  \aliascntresetthe{#1}%
}

\theoremstyle{plain}
\newtheorem{theorem}{Theorem}[section]
\DeclareSharedTheorem{lemma}{Lemma}
\DeclareSharedTheorem{proposition}{Proposition}
\DeclareSharedTheorem{corollary}{Corollary}
\DeclareSharedTheorem{claim}{Claim}
\DeclareSharedTheorem{fact}{Fact}
\DeclareSharedTheorem{observation}{Observation}
\DeclareSharedTheorem{conjecture}{Conjecture}
\newtheorem*{theorem*}{Theorem}
\newtheorem*{lemma*}{Lemma}
\newtheorem*{proposition*}{Proposition}
\newtheorem*{corollary*}{Corollary}

\theoremstyle{definition}
\DeclareSharedTheorem{definition}{Definition}
\DeclareSharedTheorem{notation}{Notation}
\DeclareSharedTheorem{example}{Example}
\DeclareSharedTheorem{problem}{Problem}
\DeclareSharedTheorem{question}{Question}

\theoremstyle{remark}
\DeclareSharedTheorem{remark}{Remark}
\DeclareSharedTheorem{note}{Note}

\crefname{theorem}{Theorem}{Theorems}
\crefname{lemma}{Lemma}{Lemmas}
\crefname{proposition}{Proposition}{Propositions}
\crefname{corollary}{Corollary}{Corollaries}
\crefname{claim}{Claim}{Claims}
\crefname{fact}{Fact}{Facts}
\crefname{observation}{Observation}{Observations}
\crefname{conjecture}{Conjecture}{Conjectures}
\crefname{definition}{Definition}{Definitions}
\crefname{notation}{Notation}{Notations}
\crefname{example}{Example}{Examples}
\crefname{problem}{Problem}{Problems}
\crefname{question}{Question}{Questions}
\crefname{remark}{Remark}{Remarks}
\crefname{note}{Note}{Notes}
\theoremstyle{plain}

\ifdefined\MathNotationLoaded\expandafter \fi
\newcommand{\MathNotationLoaded}{1.0.1}

\newcommand{\mbb}[1]{\mathbb{#1}}

\DeclarePairedDelimiter{\set}{\lbrace}{\rbrace}
\DeclarePairedDelimiter{\card}{\lvert}{\rvert}

\newcommand\restr[2]{{
  \left.\kern-\nulldelimiterspace 
  #1 
  \vphantom{\big|} 
  \right|_{#2} 
  }}

\ifdefined\ComplexityNotationLoaded\expandafter \fi
\newcommand{\ComplexityNotationLoaded}{1.0.0}

\DeclareRobustCommand{\class}[1]{\ensuremath{\mathbf{#1}}}
\DeclareRobustCommand{\lang}[1]{\ifmmode\text{\textsc{#1}}\else\textsc{#1}\fi}
\DeclareRobustCommand{\func}[1]{\ensuremath{\mathtt{#1}}}
\newcommand{\newclass}[2]{\newcommand{#1}{\class{#2}}}
\newcommand{\newlang}[2]{\newcommand{#1}{\lang{#2}}}

\newclass{\Pclass}{P}
\newclass{\Lclass}{L}
\newclass{\NP}{NP}
\newclass{\coNP}{coNP}
\newclass{\BPP}{BPP}
\newclass{\RP}{RP}
\newclass{\ZPP}{ZPP}
\newclass{\PSPACE}{PSPACE}
\newclass{\EXP}{EXP}
\newclass{\NEXP}{NEXP}
\newclass{\AC}{AC}
\newclass{\ACC}{ACC}
\newclass{\NC}{NC}
\newclass{\TC}{TC}
\newclass{\PH}{PH}
\newclass{\DTIME}{DTIME}
\newclass{\NTIME}{NTIME}
\newclass{\PromiseBPP}{PromiseBPP}
\newcommand{\poly}{\func{poly}}

\newlang{\SAT}{SAT}

\newclass{\OPP}{OPP}
\newclass{\OP}{OP}
\newclass{\BPEXP}{BPEXP}
\newclass{\coNTIME}{coNTIME}

\newlang{\OV}{OV}
\newlang{\LCS}{LCS}
\newlang{\HS}{HS}
\newlang{\maxtwosat}{Max-2-Sat}
\newlang{\maxcut}{MAXCUT}
\newlang{\perm}{Permanent}
\newlang{\minimumHammingDistance}{Minimum Hamming Distance}
\newlang{\formulaSat}{Formula Satisfiability}
\newlang{\satisfiability}{Satisfiability}
\newlang{\uniquesat}{unique-sat}
\newlang{\csp}{Csp}
\newlang{\maxsat}{MaxSat}
\newlang{\maxflow}{MaxFlow}
\newlang{\maximumMatching}{Maximum Matching}
\newlang{\cnfsat}{Cnf-Sat}
\newlang{\cktsat}{Circuit Sat}
\newlang{\turingsat}{Turing Sat}
\newlang{\ind}{Independent Set}
\newlang{\maxind}{Max Independent Set}
\newlang{\subiso}{Subgraph Isomorphism}
\newlang{\hamp}{Hamiltonian Path}
\newlang{\factoring}{Factoring}
\newlang{\graphIsomorphism}{Graph Isomorphism}
\newlang{\primality}{Primality}
\newlang{\parityfunction}{Parity}
\newlang{\hamc}{Hamiltonian Cycle}
\newlang{\CliqueProblem}{Clique}
\newlang{\colorability}{Colorability}
\newlang{\setsplitting}{Set Splitting}
\newlang{\hittingSet}{Hitting Set}
\newlang{\vertexcover}{Vertex Cover}
\newlang{\independentset}{Independent Set}
\newlang{\feedbackvertexset}{Feedback Vertex Set}
\newlang{\longestpath}{Longest Path}
\newlang{\dominatingset}{Dominating Set}
\newlang{\listcoloring}{List Coloring}
\newlang{\chromaticnumber}{Chromatic Number}
\newlang{\setcover}{Set Cover}
\newlang{\lcs}{Longest Common Subsequence}

\newlang{\subsetChain}{Maximum Length Chain of Subsets}
\newlang{\apsp}{All-Pairs Shortest Paths}
\newlang{\editDistance}{Edit Distance}
\newlang{\orthogonalVectors}{Orthogonal Vectors}
\newlang{\vectordomination}{Vector Domination}
\newlang{\frechetdistance}{Fr\'{e}chet Distance}

\DeclareRobustCommand{\enumballsat}[2]{\ifmmode\text{\textsc{Enum}}(#1,#2)\else\textsc{Enum}(\ensuremath{#1}, \ensuremath{#2})\fi}
\DeclareRobustCommand{\enumz}[1]{\ifmmode\text{\textsc{Enumz}}(#1)\else\textsc{Enumz}(\ensuremath{#1})\fi}

\DeclareRobustCommand{\enumballsatnae}[2]{\ifmmode\text{\textsc{NAE-Enum}}(#1,#2)\else\textsc{NAE-Enum}(\ensuremath{#1}, \ensuremath{#2})\fi}
\DeclareRobustCommand{\enumballsatmono}[2]{\ifmmode\text{\textsc{Enum}}^{+}(#1,#2)\else\textsc{Enum}\textsuperscript{+}(\ensuremath{#1}, \ensuremath{#2})\fi}

\DeclareMathOperator{\maj}{\mathsf{Maj}}
\DeclareMathOperator{\size}{Size}

\ifdefined\MajorityNotationLoaded\expandafter \fi
\newcommand{\MajorityNotationLoaded}{1.0.0}

\newcommand{\Sample}{{\mathrm{Sample}}} 

\newcommand{\transn}{\tau}  
\newcommand{\ntrans}{\theta} 
\newcommand{\entrans}{\Theta} 
\newcommand{\transversals}{{\mathcal T}} 
\DeclareRobustCommand{\turanl}{\ifmmode\text{Tur\'{a}n}\else Tur\'{a}n\fi}  
\newcommand{\minvalue}[1]{{\mathcal C}_{#1}}
\newcommand{\maxvalue}[1]{{\mathcal D}_{#1}}

\newcommand{\mintransset}{\mathscr{M}}
\newcommand{\ntrnz}{\Phi}

\newcommand{\turin}[2]{{\mathbb T}_{#1}^{(#2)}}  

\newcommand{\clique}[2]{{\mathbb K}_{#1}^{(#2)}}

\newcommand{\prop}[2]{\mathscr P(#1,#2)}

\newcommand{\refE}{\reflectbox{\;${\mathbf \exists}$}}

\newcommand{\qual}[2]{\text{$s=#1 \!\!\! \mod 2$,  #2}}

\newcommand{\primalopt}[1]{\mathscr{C}_{#1}}

\newcommand{\A}{\mathcal A}
\newcommand{\B}{\mathcal B}

\newcommand{\F}{\mathbb F}
\newcommand{\G}{\mathbb G}

\newcommand{\support}{\mathrm{supp}}

\newcommand{\supportplus}{\mathrm{supp}^{+}}
\newcommand{\supportnegative}{\mathrm{supp}^{-}}
\newcommand{\cost}{\mathrm{cost}}

\newcommand{\lift}{\mathrm{\mathcal L}}

\newcommand{\czeroI}{There exists a pair of variables that appears in at least two distinct $3$-clauses}
\newcommand{\czeroII}{There exists a variable that appears in at least three distinct $3$-clauses}

\newcommand{\czeroIV}{There exists a triangle configuration}
\newcommand{\czeroV}{There exists an \refE configuration}
\newcommand{\czeroVI}{There exists a variable that appears only in one clause}

\title{The Exact Complexity of Majority in Depth-Three Circuits with Bottom Fan-in Three}
\author{Mohit Gurumukhani\thanks{Cornell University, Ithaca, NY, USA. This work was supported in part by NSF Award CCF-2514586.
Email: \texttt{mgurumuk@cs.cornell.edu}}
\and
Daniel Kleber\thanks{Department of Computer Science and Engineering, University of California, San Diego. Partially supported by NSF grant 2212136. Email: \texttt{dkleber@ucsd.edu}}
\and
Ramamohan Paturi\thanks{Department of Computer Science and Engineering, University of California, San Diego. Partially supported by NSF grant 2212136. Email: \texttt{rpaturi@ucsd.edu}}
\and
Christopher Rosin\thanks{Constructive Codes, https://constructive.codes.  Email: \texttt{christopher.rosin@gmail.com}}
\and
Michael Saks \thanks{Email: \texttt{saks@math.rutgers.edu}} 
\and
Navid Talebanfard\thanks{University of Sheffield, Sheffield, UK. Supported in part by a grant of access to OpenAI models
through the ChatGPT for Academic Researchers program. Email: \texttt{n.talebanfard@sheffield.ac.uk}}
}

\declareauthor{MS}{Mike}{blue}
\declareauthor{MG}{Mohit}{red}
\declareauthor{DK}{Daniel}{green}
\date{}

\begin{document}
\maketitle

\begin{abstract}

Lecomte and Ramakrishnan (arXiv'26) recently presented a depth-three circuit of size $2^{O(\sqrt{n})}$ that computes the Majority function, thus refuting a long held conjecture. Nevertheless, as a fundamental computational primitive which ubiquitously appears across computer science, determining the \emph{exact} complexity of Majority down to the constant in the exponent, remains highly relevant. Gurumukhani et al. (CCC'24) initiated the study of optimal $\Sigma^k_3$ circuits, OR-AND-OR circuits with bottom fan-in $k$, computing the Majority function, and showed a non-trivial lower bound for $k = 3$. In this work, we determine the exact $\Sigma^3_3$ complexity of Majority up to polynomial factors. We prove our result by showing that for any $t \le n/2$, given an $n$-variate $t$-threshold 3-CNF, i.e., with no satisfying assignment with Hamming weight less than $t$, the number of weight-$t$ satisfying assignments is at most $2^t(3/2)^{(n-t)/2}$. Along with a matching construction for $t = n/2$, the result implies that the size of the smallest $\Sigma^3_3$ circuit computing Majority is exactly $\poly(n)(2/6^{1/4})^n$. Furthermore, since the matching construction is monotone, our result shows that monotone $\Sigma^3_3$ circuits are as powerful as non-monotone in computing Majority. The proof involves two main ingredients:

\begin{enumerate}
	\item \emph{Reduction to monotone.} We formulate a linear program (LP) which is defined only with respect to the monotone part of the formula. We show that bounding the cost function of this LP implies our desired upper bound on the number of minimum weight solutions of the general formula.

	\item \emph{Structure of extremal formulas.} To bound the cost function of the LP, we develop an extensive inductive argument. The inductive steps are determined by a close inspection of the local structure of extremal formulas with respect to minimum weight.
\end{enumerate}

We believe that these techniques, particularly the latter, are powerful and can be adapted to determine the exact complexity of other functions.

Our result has a natural hypergraph interpretation: it bounds the number of independent sets of  size $n-t$ under  the condition that the independence number is $n-t$. This is a generalized hypergraph Tur{\'a}n problem and coincides with the classical hypergraph Tur{\'a}n problem at threshold $t=n-k$ for $k$-uniform hypergraphs.

\end{abstract}

\newpage
\section{Introduction}

Bounded-depth circuits are one of the very few models of computation for which strong lower bounds are known. Ajtai \cite{Ajtai83} and Furst, Saxe, and Sipser \cite{FurstSS84} showed that Parity cannot be computed by constant-depth circuits of polynomial size. H\aa stad's switching lemma \cite{Hastad86} subsequently gave the tight exponential scale: for every fixed depth $d\ge2$, circuits computing Parity require size $2^{\Omega(n^{1/(d-1)})}$. Matching constructions show that this bound is optimal for Parity. Until very recently, obtaining a lower bound of $2^{\omega(n^{1/(d-1)})}$ for any fixed depth $d \ge 3$ for a function computable in polynomial time was a longstanding open problem. In a recent breakthrough, OpenAI \cite{OAI:depth-3} constructed a polynomial-time computable function that requires $2^{\omega(\sqrt n)}$-size depth-three circuits. The hard function combines linear hashing and polynomial evaluation; this construction is tailored to exhibit the required hardness. Shortly afterwards, these lower bounds were extended to arbitrary depth $d$ by Chukhin \cite{chek:depth-d} and Li and Zhong \cite{li2026shortpointlistscnfs}.

Alongside this progress, a central goal in theoretical computer science is to understand the complexity of familiar \emph{computational primitives}. We use this term informally for basic operations that recur as building blocks in more elaborate computations. Parity, for example, computes a sum modulo two and underlies binary arithmetic. Majority performs a threshold decision and ubiquitously appears across computer science. The simple descriptions and the recurring use of these operations motivate studying them in their own right.

Majority as a central computational primitive had long been held as a natural candidate to beat the depth-three square-root barrier \cite{HastadJP95}. Lecomte and Ramakrishnan \cite{LecomteRamakrishnan2026} have now shown that, for every fixed $d\ge2$, \emph{every symmetric Boolean function} has depth-$d$ circuits of size $2^{O(n^{1/(d-1)})}$. For Majority, this matches the known lower bound in its asymptotic exponent; in particular, its unrestricted depth-three complexity is $2^{\Theta(\sqrt n)}$. Nevertheless, we argue that determining the \emph{exact} complexity of Majority remains highly relevant. To illustrate this point, note that in the general circuit model, i.e., circuits with fan-in two of unrestricted depth over an arbitrary basis, the exact complexity of Majority is open and is only known to be between $2.5n - o(n)$ and $4.5n + o(n)$ (see \cite{DemenkovKKY10}). Furthermore, Gurumukhani et al. \cite{GurumukhaniPaturiPudlakSaksTalebanfard_2024_CCC} showed that determining the exact depth-three complexity of Majority has algorithmic consequences.


\begin{problem}
Determine the smallest $c > 0$ such that Majority can be computed by depth-three circuits of size $2^{c\sqrt{n}}$.
\end{problem}

\paragraph*{Depth-three circuits and Valiant's program.} The study of depth-three circuits was advocated by Valiant \cite{Valiant77} who showed that any bounded-fan-in, linear-size circuit of logarithmic depth can be computed by a depth-three circuit of size $2^{O(n/\log\log n)}$. Thus, sufficiently strong depth-three lower bounds imply superlinear lower bounds for circuits of logarithmic depth. A particularly interesting class is $\Sigma_3^k$: OR-AND-OR circuits whose bottom gates have fan-in at most $k$, that is, disjunctions of $k$-CNFs. As a natural measure of size, we denote the minimum top fan-in of a $\Sigma_3^k$ circuit computing $f$ by $\size_3^k(f)$, and by $\size_3^{+k}(f)$ when the circuit is monotone. For fixed $k$, top fan-in and total gate count differ by at most a polynomial factor, because a $k$-CNF has only polynomially many distinct clauses. Golovnev, Kulikov, and Williams \cite{GKW21circuit} revisited Valiant's approach and showed that even a fixed bottom fan-in class captures unrestricted circuits: a near-maximal $2^{n-o(n)}$ lower bound for $\Sigma_3^{16}$ would imply new unrestricted circuit lower bounds. This gives a strong motivation for studying $\Sigma^k_3$ circuits for small values of $k$.


\paragraph*{The bounded bottom fan-in lower bound frontier.} Paturi, Pudl{\'a}k, and Zane \cite{PaturiPZ99} (PPZ) showed that Parity requires $\Sigma_3^k$ circuits of size $\Omega(2^{n/k})$. Their argument also gives the tight unrestricted depth-three lower bound $\Omega(n^{1/4}2^{\sqrt n})$ for Parity. Paturi, Pudl{\'a}k, Saks, and Zane \cite{PaturiPSZ05} (PPSZ) extended these techniques to recognizing codewords of suitable error-correcting codes, including BCH codes, improving the constant multiplying $n/k$ in the exponent. Improving this constant remains a central challenge. We note that the OpenAI lower bound seems to give a constant substantially worse than even PPZ.




To determine the depth-three complexity of Majority, Gurumukhani et al.\ \cite{GurumukhaniPaturiPudlakSaksTalebanfard_2024_CCC,GurumukhaniKunnemannnPaturi_2024_arxiv} proposed the following problem.

\begin{problem}
For $k \ge 3$, determine $\size_3^k(\maj)$.
\end{problem}

Originally they conjectured a $2^{\Omega(n\log k/k)}$ lower bound. Together with a bottom fan-in reduction, the conjectured bound would give an unrestricted $2^{\Omega(\sqrt{n\log n})}$ lower bound. Lecomte, Ramakrishnan, and Tan \cite{LecomteRT22} proved a special case of this conjecture.

The recent construction of \cite{LecomteRamakrishnan2026} refutes the conjecture. It is not hard to see that their construction gives $\size_3^k(\maj) \le 2^{5n/k}$. It should be noted that the construction does not yield anything for very small values of $k$ such as $k = 3$. It is not clear if the constant in the exponent can be improved. Therefore, proving optimal lower bounds for Majority for fixed bottom fan-in $k$, may result in beating the lower bound of PPSZ. Furthermore, since the construction is highly non-monotone, it remains open to determine the power of monotone depth-three circuits in computing Majority. We hesitate to state this as a conjecture, but it is conceivable that $\size_3^{+k}(\maj) \ge 2^{\Omega(n\log k/k)}$.

\paragraph*{Exact bounds and satisfiability algorithms.} The study of $\Sigma^k_3$ circuits involves studying properties of $k$-CNFs and their satisfying assignments that may be useful in developing $k$-SAT algorithms. Indeed, the core technique in the PPZ and PPSZ lower bounds is an instantiation of this intuition. An algorithm solving $k$-SAT is nontrivial if it has running time $2^{(1-\epsilon_k)n}$ for some $\epsilon_k>0$, which we call its \emph{savings}. PPZ and PPSZ used their combinatorial arguments to give such algorithms; subsequent analyses have improved the bounds \cite{Scheder21,JiangCai2026}. Gurumukhani et al. \cite{GurumukhaniPaturiPudlakSaksTalebanfard_2024_CCC} showed that a natural approach towards lower bounds for Majority also gives nontrivial $k$-SAT algorithms. However, OpenAI's recent refutation of ETH seems to obviate the need for such an approach to $k$-SAT.

\paragraph*{Reflections on the claimed refutation of ETH.} OpenAI \cite{OAI:ETH} has recently published a sophisticated algebraic algorithm claiming to solve 3-SAT in subexponential time. The algorithm seemingly goes against the above intuition and ignores the structure and properties of the input formula and its satisfying assignments. As the correctness of the result continues to be verified, we should seriously consider the possibility that understanding the combinatorics of satisfying assignments may not be a necessary step for efficient SAT algorithms. However, as matters stand, we believe that understanding the representational limits of a fundamental computational model (in this case $k$-CNFs) and their algorithmic consequences is a justified direction to explore, even though the implied algorithms need not be more efficient than other algorithms solving their corresponding tasks.


\medskip
Gurumukhani et al. \cite{GurumukhaniPaturiPudlakSaksTalebanfard_2024_CCC} captured the connection between Majority and $k$-SAT by defining the local enumeration problem.

\paragraph*{Local enumeration.}
\enumballsat{k}{t} is defined as follows: for a natural number $k$ and an integer $t$, given an $n$-variate $k$-CNF $F$ with no satisfying assignment of Hamming weight less than $t$, enumerate all satisfying assignments of Hamming weight exactly $t$.


The following lemma is immediate.

\begin{lemma}[\cite{GurumukhaniPaturiPudlakSaksTalebanfard_2024_CCC}]
\label{thm: local enum implies algorithm and maj lb}
Suppose that, for each integer $0\le t\le\lfloor n/2\rfloor$, \enumballsat{k}{t} can be solved by an algorithm in expected time $b(n,k,t)$. Then, for even $n$,
\[
 \size_3^k(\maj_n)\ge\frac{\binom{n}{n/2}}{b(n,k,n/2)}.
\]
Moreover, $k$-SAT can be solved in expected time
$O\bigl(\sum_{t=0}^{\lfloor n/2\rfloor}b(n,k,t)\bigr)$.
\end{lemma}



Gurumukhani et al.\ \cite{GurumukhaniPaturiPudlakSaksTalebanfard_2024_CCC} considered $k=3$ as the first nontrivial instance and solved \enumballsat{3}{n/2} in expected time $1.598^n$. Their subsequent work \cite{GurumukhaniP0T25} obtains the optimal $\poly(n)6^{n/4}$ bound for enumerating only the Not-All-Equal solutions under the same promise.

Local enumeration has a natural combinatorial interpretation.


\paragraph*{The generalized hypergraph \texorpdfstring{\turanl}{Turán} problem.}
The \turanl\ problem for $k$-uniform hypergraphs can be stated in terms of monotone $k$-CNFs: what is the minimum number of clauses in an $n$-variate uniform monotone $k$-CNF for which every satisfying assignment has weight at least $t$? Our problem instead asks to maximize the number of satisfying assignments of weight exactly $t$ under the same threshold condition. There is a precise generalized \turanl\ interpretation (see \cite{ALON2016146} for more detail on generalized \turanl\ problems). Put $m=n-t$ and take the edge-complement $J$ of the clause hypergraph. A weight-$t$ transversal corresponds to an independent $m$-set in the clause hypergraph, and hence to a complete $k$-uniform hypergraph on $m$ vertices in $J$. The threshold condition says that $J$ contains no complete $k$-uniform hypergraph on $m+1$ vertices. Thus we want to maximize the number of copies of $K_m^{(k)}$ in a $K_{m+1}^{(k)}$-free $k$-graph, for $m\ge k$.

At $t=n-k$, this is exactly the usual problem of maximizing the number of edges in a $K_{k+1}^{(k)}$-free $k$-graph \cite{GurumukhaniPaturiPudlakSaksTalebanfard_2024_CCC}.
For $k=3$, it is the tetrahedron problem \cite{Sidorenko_1995_gc,Keevash_2011}.
Our problem is therefore a natural generalization of the hypergraph \turanl\ problem, with exact results in the threshold range $t\le n/2$.
The small \turanl-style configurations that enter our extremal constructions explain this connection concretely; we do not claim that the present range resolves the tetrahedron problem.

\subsection{Our Results}

We determine the exact $\Sigma^3_3$ complexity of Majority up to a polynomial factor. Furthermore, we show that for bottom fan-in three, the monotone and non-monotone complexity of Majority coincide. Given a CNF $F$, by $\ntrans_t(F)$ we denote the number of satisfying assignments of $F$ of Hamming weight exactly $t$. A CNF is $t$-\emph{threshold} if it has no satisfying assignment of weight less than $t$.

\begin{theorem}[Threshold counting for $3$-CNFs]\label{thm:main}
Let $0\le t\le n$, and let $F$ be a $t$-threshold $3$-CNF on $n$ variables. Then
\[
 \theta_t(F)\le2^t\left(\frac32\right)^{(n-t)/2}.
\]
In particular, $\theta_{\lfloor{n/2}\rfloor}(F)\le6^{n/4}$, with equality attainable whenever $4\mid n$.
\end{theorem}

\begin{corollary}\label{cor:intro-signed-circuit}
For all $n$,
\[
 \size_3^3(\maj_n)\ge\frac{\binom{n}{\lfloor{n/2}\rfloor}}{6^{n/4}}.
\]
For all $n$ there is a matching upper bound showing that the general and monotone circuit models have the same size:
\[
 \size_3^3(\maj_n)
 =\left(\frac{2}{6^{1/4}}\right)^{n+o(n)},
 \qquad
 \size_3^{+3}(\maj_n)
 =\left(\frac{2}{6^{1/4}}\right)^{n+o(n)}.
\]
\end{corollary}


Furthermore, our proof yields an optimal algorithm for a sampling variant of the local enumeration problem for $k = 3$.

\begin{theorem}[Local sampling algorithm]\label{thm: main generalized}
For every $0\le t\le n/2$, there is a polynomial-time randomized algorithm such that given an $n$-variate $t$-threshold 3-CNF, it outputs every weight-$t$ satisfying assignment with probability at least $2^{-t}\left(\frac23\right)^{(n-t)/2}$.
\end{theorem}


\subsection{Our Techniques}

\paragraph*{Extremal CNFs and top-down covers.}
To obtain circuit lower bounds, we use the well-known top-down argument. Consider a $\Sigma_3^k$ circuit computing a Boolean function $f:\{0,1\}^n\to\{0,1\}$. Let $s$ be the fan-in of the top OR gate. Then $f$ is an OR of $s$ many $k$-CNFs, each of which is \emph{consistent} with $f$ in that each accepts only inputs in $f^{-1}(1)$. For any nonempty set $A\subseteq f^{-1}(1)$, one of these CNFs must accept at least $|A|/s$ members of $A$. This raises the natural question of studying \emph{extremal} $k$-CNFs with respect to an appropriately chosen set $A$. For Majority, the useful set $A$ is the middle slice of the Boolean cube, rather than the entire set $f^{-1}(1)$.

Variable permutations act transitively on each Hamming layer. Consequently, translates of a CNF consistent with Majority accepting an optimal number of solutions in the middle slice can efficiently cover that layer. For monotone formulas, covering the boundary layer also covers all higher layers; for a general symmetric function, its different accepting layers must be treated separately. This is also the viewpoint behind the new upper bound of Lecomte and Ramakrishnan \cite{LecomteRamakrishnan2026}: construct CNFs accepting many inputs of a specified weight, and combine their translates through an OR.
Their tests compare shifted weights of different blocks, rather than fixing each block's contribution independently.
Thus their upper bound and our lower bound study the same covering question from opposite directions, under different width restrictions.

We now address how we determine extremal formulas and prove their optimality.

\paragraph*{Modular extremal constructions.} Our intuition, based on optimal constructions for small $n$, suggested searching for \emph{modular} constructions: formulas that are conjunctions of small CNFs, which we call \emph{building blocks}, on disjoint sets of variables. The building blocks include complete monotone $3$-CNFs and small \turanl-style hypergraphs. Combining these blocks in an appropriate fashion gives extremal monotone constructions and concrete bounds for $t\le n/2$.

{}

\paragraph*{Optimality for monotone formulas.} We can prove the optimality by an extensive induction on $n$ and $t$. The constructions and their corresponding bounds allow us to formulate very precise inductive hypotheses. The induction requires hypotheses not only for smaller $n$ and $t$, but also for formulas containing a certain number of $2$-clauses. In the inductive step, we identify a particular substructure, simplify the formula by setting some variables, and apply the appropriate inductive bound. The catalogue of these substructures is determined by the local structure of our building blocks. For example one substructure we consider is a variable that appears in three clauses, or another is a set of two variables that appear in two clauses together. 

In this sense, upper-bound constructions guide the lower-bound proof: guessing the extremal examples is an essential step toward proving their optimality. We will include the proof in the full version of the paper.

\paragraph*{Optimality for general CNFs.} For the extension to general CNFs, an appropriate linear program (LP) allows us to essentially reduce the problem to the monotone case. To make the reduction go through, we sacrifice some amount of precision. Nevertheless, our results remain asymptotically optimal. Given a formula $\mbb{F}$ we write it as  $\mbb{F} = \mbb{G} \wedge \mbb{H}$, where $\mbb{H}$ contains all monotone clauses and each clause of $\mbb{G}$ has at most two positive literals. We regard $\mbb{H}$ as a CNF on all $n$ variables, including variables that occur only in $\mbb{G}$. For a transversal $T$ of $\mbb{H}$, i.e. a set of variables that hit every clause, fixing $T$ to one leaves the restricted formula $\mbb{G}[T = 1]$. The number of positive variables in every clause of this formula is at most two. It can then easily be argued that the number of weight-$t$ solutions of $\mbb{F}$ that extend $T$ is at most $2^{t - |T|}$. Therefore, the total number of weight-$t$ solutions of $\mbb{F}$ is at most $W_{t, \mbb{H}} := \sum_{T \in \mintransset(\mbb{H})}2^{t - |T|}$, where $\mintransset(\mbb{H})$ is the set of minimal transversals of $\mbb{H}$. Note that this quantity is expressed purely in terms of the monotone part of $\mbb{F}$. If we could somehow bound this quantity by our desired target upper bound, then we would be done. Remarkably, our monotone proof can be extended to prove the bound for $t \le \frac{n + 2}{3}$. But it provably fails for larger $t$. This should not come as a surprise. Each weight-$t$ solution of $\mbb{F}$ may contain several minimal transversals of $\mbb{H}$ and hence $W_{t, \mbb{H}}$ overcounts the solutions.

To overcome this difficulty, we will consider a fractional version of $W_{t, \mbb{H}}$. To each transversal $T$ of $\mbb{H}$ we assign a non-negative weight $w_T$. Let $W'_{t, \mbb{H}}$ be the minimum of $\sum_{T \in \mintransset(\mbb{H})}w_T 2^{t - |T|}$, subject to $\sum_{T \in \mintransset(\mbb{H}): T \subseteq S} w_T \ge 1$ for every transversal $S$ of size $t$. It is easy to see that $W'_{t, \mbb{H}}$ is an upper bound on the number of weight-$t$ solutions of $\mbb{F}$.

For carefully chosen parameters $p=\sqrt{3/2}$ and $r=p/2$, we will work with essentially a rescaled version of this quantity. We want to minimize $\sum_T w_T r^{|T|}$ under certain constraints involving the coverage of transversals where $w_T$ is the weight assigned to the transversal $T$ of $\mbb{H}$. The details are given in Section \ref{sec:LP}. We bound the cost of this LP by $p^n$ via an inductive proof. The inductive proof follows the structure of the monotone argument. This proof also identifies a set of substructures and a set of corresponding restrictions to variables. The inductive hypothesis gives us weights for these restrictions. We then lift these weights back to the original formula. We have to do this carefully to satisfy the coverage requirement. A restriction may contribute only part of a target's required coverage. However, we require that the sum of the contributions of all restrictions should be at least 1. 


\paragraph*{The sampling algorithm.}  The algorithm is based on the following idea: the proof that shows the existence of weights that guarantee a cost of at most $p^n$ can be interpreted as an algorithm that samples each min-size transversal of size $t$ with probability at least $r^t/p^n$. The proof looks for a set of substructures and for each substructure it identifies a set of restrictions along with a scaling factor. The algorithm picks one of the restrictions with an appropriate probability, applies it, then continues recursively. The algorithm and its full analysis appear in \cref{sec:algorithmic_aspect}.

\section{CNFs and Transversals}

A \emph{CNF} (Conjunctive Normal Form) is a conjunction of
clauses, each a disjunction of literals. A literal on a Boolean
variable $x$ is either $x$ (positive) or $\neg x$ (negative).
We represent a CNF by $\mbb{F}=(V,F)$, where $V$ is the set of
Boolean variables and $F$ is the collection of clauses,
regarded as sets of literals on distinct variables.

\begin{notation}
We use $n(\mbb{F})=|V|$ and $m(\mbb{F})=|F|$.
A $k$-clause is a clause of size $k$; similarly, we use
$k$-set and $k$-edge. A $k$-CNF has clauses of size at most
$k$ and is $k$-\emph{uniform} if every clause has size $k$.
\end{notation}

\begin{definition}[$t$-threshold CNFs]
A CNF is $t$-\emph{threshold} if every satisfying assignment
has at least $t$ variables assigned $1$.
\end{definition}

A clause is \emph{monotone} if all its literals are positive,
and a CNF is monotone if all its clauses are monotone.
A monotone CNF $\mbb{F}=(V,F)$ can also be viewed as a
hypergraph or a \emph{set system}, with $V$ as its vertices
or elements and $F$ as its edges or sets.

\begin{definition}[Transversals]
A $t$-\emph{transversal} of a CNF $\mbb{F}=(V,F)$ is a
satisfying assignment with exactly $t$ variables assigned
$1$, identified with the subset of $V$ assigned $1$.
A \emph{transversal} is a $t$-transversal for some $t\geq 0$.
For a monotone CNF, a transversal is precisely a subset of
$V$ that intersects every clause in $F$.
\end{definition}


\begin{definition}[Transversal number]
The \emph{transversal number} $\transn(\mbb{F})$ of a
satisfiable CNF $\mbb{F}$ is the size of its minimum-size
transversals.
\end{definition}

\begin{notation}
We use $\transversals_t(\mbb{F})$ to denote the set of $t$-transversals of $\mbb{F}$ and $\ntrans_t(\mbb{F})$ to denote their number. Furthermore, we use $\transversals(\mbb{F}) = \bigcup_{t \ge 0} \transversals_t(\mbb{F})$. We use $\mintransset(\mbb{F})$ to denote the set of (inclusion-)minimal transversals of $\mbb{F}$.
\end{notation}

\begin{definition}[$t$-extremal $k$-CNFs]
A $t$-threshold $k$-CNF $\mbb{F}$ is $t$-\emph{extremal}
if $\ntrans_t(\mbb{F})$ is maximum among all $t$-threshold
$k$-CNFs on $n(\mbb{F})$ variables.
For $n,k\geq 1$ and $0\leq t\leq n$, we denote this
maximum by $\entrans(n,t,k)$.
\end{definition}

For extremal monotone $k$-CNFs, it suffices to consider
uniform CNFs, possibly of width less than $k$.

\begin{fact}[Uniformity \cite{GurumukhaniPaturiPudlakSaksTalebanfard_2024_CCC}]
For $1\leq k\leq n$ and $1\leq t\leq n-k$, every
$t$-threshold $k$-CNF has a $t$-threshold 
$k$-uniform CNF with the same set of $t$-transversals.
Moreover, every
$t$-threshold monotone
$k$-CNF has a $t$-threshold monotone
$k$-uniform CNF with the same set of $t$-transversals.
For $1\leq i\leq k$, the monotone CNF consisting of all
$i$-clauses is $(n-i+1)$-extremal.
\end{fact}

\subsection{Circuits and Transversals}

A $\Sigma_3^k$ circuit is a depth-3 circuit with an
$\mathbf{OR}$ gate at the top and $k$-CNFs as its depth-2
sub-circuits. Its \emph{size} is the number of these
sub-circuits. A monotone $\Sigma_3^k$ circuit has only
monotone $k$-CNFs as its depth-2 sub-circuits.

\begin{definition}
For $0\leq t\leq n$, the $t$-\emph{threshold function}
$f$ on $n$ variables satisfies
$f(x_1,\ldots,x_n)=1$ if and only if
$\sum_{i=1}^n x_i\geq t$.
\end{definition}

Every depth-2 sub-circuit of a $\Sigma_3^k$ circuit
computing the $t$-threshold function is a $t$-threshold
$k$-CNF. Since each has at most $\entrans(n,t,k)$
$t$-transversals, we obtain the following lower bound.

\begin{fact}
\label{fact: circuit lb from extremal}
Every $\Sigma_3^k$ circuit computing the $t$-threshold
function has size at least
\[
\frac{\binom{n}{t}}{\entrans(n,t,k)}.
\]
\end{fact}

As a partial converse, we have 
for any $t$-threshold monotone $k$-CNF
$\mbb{F}$ with $\ntrans_t(\mbb{F})>0$, permutations of its
variables yield a monotone $\Sigma_3^k$ circuit computing
the $t$-threshold function with size at most
$\poly(n)\binom{n}{t}/\ntrans_t(\mbb{F})$
\cite{GurumukhaniKunnemannnPaturi_2024_arxiv}.
In particular, we have the following.

\begin{fact}
\label{fact: extremal to circuit}
If a $t$-extremal $k$-CNF is monotone,
then there is
a monotone $\Sigma_3^k$ circuit computing the $t$-threshold
function whose size is at most
\[
\poly(n)\frac{\binom{n}{t}}{\entrans(n,t,k)}.
\]
\end{fact}

\section{Constructions for Extremal Threshold CNFs}
In this section, we construct
$k$-CNFs over $n\geq 1$ variables for transversal number $0\leq t\leq \lfloor \frac{n}{2}\rfloor$. In the following sections, we argue that they are  $t$-extremal.
We first consider certain constructions for bounded $n$ which will serve as  
\emph{building blocks} for general constructions.
The basic building blocks come in two flavors: 
cliques and \turanl-style constructions.

\subsection{Cliques}

\begin{definition}[Clique]
For $\ell\geq k\geq 1$, let $\clique{\ell}{k}$ denote the monotone $k$-CNF of all monotone $k$-clauses on $\ell$ variables.
\end{definition}
The following two cliques are widely used in our constructions:
$\clique{3}{3}  = \{\{1,2,3\}\}$, a single clause on three variables, is a $1$-threshold monotone $3$-CNF where $\ntrans_1(\clique{3}{3}) = 3$; 
$\clique{4}{3}  = \{\{1,2,3\},\{1,2,4\},\{1,3,4\},\{2,3,4\}\}$ is a $2$-threshold monotone $3$-CNF, consisting of all width $3$ monotone clauses over  $4$ variables, where $\ntrans_2(\clique{4}{3}) =6$.
More generally, we can easily show the following fact regarding cliques:

\begin{fact}
$\clique{\ell}{k}$ is a $(\ell-k+1)$-extremal  $k$-uniform monotone CNF on $\ell$ variables where the number of variables $n(\clique{\ell}{k}) = \ell,$ the number of edges $m(\clique{\ell}{k}) = \binom{\ell}{k},$ the transversal number $\tau(\clique{\ell}{k}) = \ell-k+1,$  
and the number of transversals of size exactly $\ell - k + 1$ is $\entrans(\ell,\ell-k+1,k) = \ntrans_{\ell-k+1}(\clique{\ell}{k}) = \binom{\ell}{\ell-k+1}.$
\end{fact}

To see this fact, notice that any set of $\ell-k$ variables cannot form a transversal since there is an edge over the missing $k$ variables. Similarly, any set of $\ell-k+1$ variables indeed forms a transversal since there is no edge solely containing the missing $k-1$ variables.
Lastly, this is extremal as maximum number of transversals of size $\ell-k+1$ is trivially upper bounded by $\binom{\ell}{\ell-k+1}$, and the clique construction attains this bound.

\subsection{\texorpdfstring{\turanl-style constructions}{Turán-style constructions}}

Let $n \geq 3$. We observe the problem of determining the maximum number of $t$-transversals for monotone $3$-CNF  is closely related to the hypergraph Tur{\'a}n problem. In particular, $t = n - 3$ exactly captures the tetrahedron problem \cite{Sidorenko_1995_gc, Keevash_2011}. Consider the following $(n-3)$-threshold formula $\turin{n}{3}$ which is obtained by what we call a $\turanl$-style construction. Partition the $n$ variables into three parts where the parts have sizes as equal as possible and organize them in a circular fashion. $\turin{n}{3}$ consists of the following $3$-clauses:

\begin{enumerate}
\item For each part, all 3-clauses of the part
\item For each part, all 3-clauses of the form where two variables come from the part and one variable from the adjacent part in the circular order.
\end{enumerate}

We use the following \turanl-style building blocks in our constructions.

\begin{example}
$\turin{5}{3}  = \{\{1,2,3\},\{3,4,5\},\{1,2,4\}\}$  is a $2$-threshold $3$-uniform monotone CNF, 
$n(\turin{5}{3})   = 5$,
$m(\turin{5}{3})   = 3$,
$\transn(\turin{5}{3})   = 2$, and
$\ntrans_2(\turin{5}{3})   = 7$.
\end{example}

\begin{example}
$\turin{6}{3}
= \{ 
\{x_0, y_0, x_1\}, \{x_0, y_0, y_1\},
\{x_1, y_1, x_2\}, \{x_1, y_1, y_2\},
\{x_2, y_2, x_0\}, \{x_2, y_2, y_0\}\}$ is a $3$-threshold $3$-uniform set system, 
$n(\turin{6}{3}) = 6$,
$m(\turin{6}{3}) = 6$,
$\tau(\turin{6}{3}) = 3$, \text{ and }
$\ntrans_3(\turin{6}{3}) = 14$.
\end{example}

For these parameters, observe that every single edge of width $3$ exactly rules out a single transversal at threshold $t$. Hence, maximizing the number of transversals is exactly the same as minimizing the number of edges required to attain the threshold. Constructions minimizing the latter quantity are precisely \turanl\, constructions.

\begin{fact}
\begin{align*}
\entrans(5,2,3) &= \ntrans_2(\turin{5}{3}) = 7\\
\entrans(6,3,3) &= \ntrans_3(\turin{6}{3}) = 14
\end{align*}
\end{fact}

\subsection{Constructions for $n\geq 1, k=3$ and $0\leq t\leq \lfloor \frac{n}{2} \rfloor$}
Our constructions for $n\geq 1$ and $0\leq t\leq \lfloor \frac{n}{2}\rfloor$ are limited to the case $k=3$. They use $\clique{3}{3}$, $\clique{4}{3}$, $\turin{5}{3}$, and
$\turin{6}{3}$ as building blocks. Our constructions vary based on the parity of $3t-n$, so the following reparameterization is useful.

\begin{definition}
For $k=3$, we  rewrite $n=3t-s$ where $s$ is called the \emph{deficit} of the variables (elements or vertices) with respect to the threshold $t$.  Our constructions and bounds depend on the parity of $s$. To emphasize the role of $s$, we specialize $\entrans(n,t,k)$
for $k=3$ and use the function $\ntrnz_0(s,t)$ to denote the maximum number of $t$-transversals of a $t$-threshold monotone $3$-CNF  with deficit $s$.\footnote{Note that $s$ can be negative (when $t < n/3$) or $0$ (when $t = n/3$) or positive (when $t > n/3$).} In other words, $\ntrnz_0(s,t) = \entrans(3t-s,t,3)$.
\end{definition}

We need the following notation to represent larger CNFs composed of smaller CNFs. If $\mbb{F}_1 = (V_1,F_1)$ and $\mbb{F}_2 = (V_2,F_2)$ are monotone $k$-CNFs over disjoint sets of variables,
the sum $\mbb{F}_1 +\mbb{F}_2$ of $\mbb{F}_1$ and $\mbb{F}_2$
represents the monotone $k$-CNF $\mbb{F} = (V_1\cup V_2, F_1\cup F_2)$. For a monotone $k$-CNF $\mbb{F}$, $\ell \cdot \mbb{F}$ represents the monotone $k$-CNF obtained by summing $\ell$ disjoint copies of $\mbb{F}$.


Let $s\leq t$ and $t\geq 0$ (equivalently $n\geq 2t\geq 0$).

Let $\mbb{P}^3_{s,t}$ be the monotone $3$-CNF over $n=3t-s$ variables defined as follows.

\begin{align*}
\mbb{P}^3_{s,t} &=
\begin{cases}
t \cdot  \clique{3}{3} &\text{$s\leq 0$, $t\geq 0$}\\
(t-2)  \cdot \clique{3}{3} + 1  \cdot \turin{5}{3} &\text{$s=1$, $t\geq 1$}\\
(t-s)  \cdot \clique{3}{3} + \frac{s}{2}  \cdot \clique{4}{3} &\qual{0}{$2\leq s \leq t$}\\
(t-s)  \cdot \clique{3}{3} + \frac{s-3}{2} \cdot  \clique{4}{3} + 1\cdot \turin{6}{3} &\qual{1}{$2\leq s \leq t$}\\
\end{cases}
\end{align*}

\begin{fact}
$\mbb{P}^3_{s,t}$ is a $t$-threshold $3$-uniform monotone CNF.
\end{fact}

\begin{fact} \label{fact:cons}
\begin{align*}
\ntrans_t(\mbb{P}^3_{s,t}) &=
\begin{cases}
3^t  &\text{$s\leq 0$, $t\geq 0$}\\
(\frac{2}{3})^{\frac{s}{2}} 3^t &\qual{0}{$1\leq s \leq t$}\\
\frac{7}{9} (\frac{2}{3})^{\frac{s-1}{2}} 3^t &\qual{1}{$1\leq s \leq t$}\\
\end{cases}
\end{align*}
\end{fact}

\begin{remark}
$ \clique{3}{3} + \turin{6}{3}$ can be replaced with $\clique{4}{3} + \turin{5}{3}$ without affecting the number of minimum-size transversals.
\end{remark}

We summarize the results of this section by the following theorem. It provides lower bounds for the maximum number of minimum-size transversals $\ntrnz_0(s,t)$ of monotone $3$-CNFs over $n=3t-s$ elements with transversal number and deficit satisfying the conditions $s\leq t$ and $t\geq 0$.

\begin{theorem}
\label{thm: construct extremal k cnf}
\begin{align*}
\ntrnz_0(s,t) &=
\begin{cases}
3^t  &\text{$s\leq 0$, $t\geq 0$}  \\
\binom{3t-s}{t}  &\text{$0\leq 2t-2\leq s\leq 2t$, $t\geq 1$} \\
\end{cases}\\
\ntrnz_0(s,t) &\geq
\begin{cases}
(\frac{2}{3})^{\frac{s}{2}} 3^t  &\qual{0}{$1\leq s \leq t$}\\
\frac{7}{9} (\frac{2}{3})^{\frac{s-1}{2}} 3^t &\qual{1}{$1\leq s \leq t$}\\
\end{cases}
\end{align*}
\end{theorem}

\begin{proof}
The first equation  follows from \cite{GurumukhaniKunnemannnPaturi_2024_arxiv} and its main result is reproduced below.
The second equation  is straightforward and we state as a fact below.
The remaining two inequalities follow from Fact \ref{fact:cons}.  
\end{proof}

Observe that when $s = t$ and $s$ is even, we are in the setting of $t = n/2$ and then, $\ntrnz_0(t,t) \ge 6^{t/2} = 6^{n/4}$. In the next section, we prove a matching upper bound on $\ntrnz_0(s, t)$ and for the setting of $s = t$, we will exactly obtain \cref{thm:main}.

\begin{theorem}[\cite{GurumukhaniKunnemannnPaturi_2024_arxiv}]
For $k,n\geq 1$ and $0\leq t\leq \lfloor\frac{n}{k}\rfloor$, 
$\mbb{S}^k_{t}= t \cdot  \clique{k}{k}$ is the unique (up to isomorphism) $t$-extremal monotone $k$-CNF on $n$ variables.
\end{theorem} 

For $t = n-k+i$  where $1\leq i\leq k$, we have

\begin{fact}
The $(k-i+1)$-uniform monotone CNF consisting of all $(k-i+1)$-clauses is a $t\,(= n-k+i)$-extremal monotone $k$-CNF and $\entrans(n,t,k) = \binom{n}{t}$.
\end{fact}

\section{LP formulation for bounding the number of transversals}
\label{sec:LP}

In this section we develop the technical tools needed for bounding the number of transversals. More specifically, given a $t$-threshold CNF $\F$, we formulate an LP only based on the monotone part of $\F$ such that an upper bound on its cost function can be used to bound the number of $t$-transversals of $\mbb{F}$. Therefore, in a sense, our proof is a reduction from non-monotone to monotone formulas. 

Let $\F = \G\land \mbb{H}$ be a $t$-threshold $3$-CNF on the set $V$ of variables, where $\mbb{H}$ consists of all positive clauses. Every clause of $\G$ has at most two positive literals. If $T\in \transversals(\mbb{H})$, setting every
variable in $T$ to one satisfies $\mbb{H}$ and leaves
$\F[T=1]=\G[T=1]$. We have
\begin{lemma}\label{lem:restriction-weight} For every $T\in \transversals(\mbb{H})$,
\begin{align*}
\sum_{\substack{U\in \mintransset(\F)\\T\subseteq U}}  2^{-\vert U\setminus T\vert}\leq 1 .
\end{align*}
\end{lemma}
\begin{proof}
First, restrictions of this form preserve minimality. If
$U\in\mintransset(\F)$ and $T\subseteq U$, then
$U\setminus T\in\mintransset(\F[T=1])$: any proper satisfying subset of
$U\setminus T$ would, after adjoining $T$, give a proper satisfying
subset of $U$.

We next prove that every CNF $\G$ with at most two positive literals per
clause satisfies
\[
\sum_{U\in\mintransset(\G)}2^{-|U|}\le1.
\tag{*}\label{eq:positive-width}
\]
Induct on the number of variables. If $\G$ is unsatisfiable, the sum is
zero. If the empty set satisfies $\G$, it is the unique minimal satisfying
set, and the sum is one. Otherwise, some clause $C$ is false under the
all-zero assignment. Since the unsatisfiable case has been excluded, $C$
is a nonempty all-positive clause with $|C|\le2$.
Every $U\in\mintransset(\G)$ meets $C$. For each $x\in U\cap C$, the preceding
minimality argument gives
$U\setminus\{x\}\in\mintransset(\G[x=1])$. For a fixed $x$ this map is injective.
The residual formula still has at most two positive literals per clause
and has one fewer variable. Therefore
\[
\begin{aligned}
\sum_{U\in\mintransset(\G)}2^{-|U|}
&\le \sum_{x\in C}\sum_{\substack{U\in\mintransset(\G)\\x\in U}}2^{-|U|}\\
&\le \frac12\sum_{x\in C}
       \sum_{S\in\mintransset(\G[x=1])}2^{-|S|}
\le \frac{|C|}{2}\le1.
\end{aligned}
\]
Finally, $U\mapsto U\setminus T$ injects the minimal satisfying sets of
$\F$ containing $T$ into $\mintransset(\G[T=1])$. Applying~\eqref{eq:positive-width}
to $\G[T=1]$ proves the lemma.
\end{proof}

For $r>0$, define
\[
Z(\F,r)=\sum_{U\in\mintransset(\F)}r^{|U|}.
\]

\begin{fact}\label{fact:count-weight}
If $\F$ is $t$-threshold, then $\theta_t(\F)\le r^{-t}Z(\F,r)$.
\end{fact}
\begin{proof}
Every satisfying set of size $t$ is minimal, since a proper
satisfying subset would have size less than $t$. Each contributes $r^t$
to $Z(\F,r)$, and all other contributions are nonnegative.
\end{proof}


\subsection{A packing relaxation and its covering dual}

Fact~\ref{fact:count-weight} reduces the counting problem to bounding
$Z(\F,r)$. Lemma~\ref{lem:restriction-weight} gives a family of constraints
on the sets contributing to this sum: for each $T\in\transversals(\mbb{H})$, the
total contribution of minimal transversals $U$  of $\F$ that include $T$, discounted by $2^{-|U\setminus T|}$, is at
most one. We use all these constraints simultaneously.

Introduce a nonnegative variable $f^*(U)$ for each $U\in\mintransset(\F)$, and
maximize the weighted sum subject to the inequalities supplied by the
lemma:
\begin{equation}
\boxed{
\begin{aligned}
\max\quad &\sum_{U\in\mintransset(\F)}f^*(U)r^{|U|}\\
\text{subject to}\quad
&\sum_{\substack{U\in\mintransset(\F)\\T\subseteq U}}
 f^*(U)2^{-|U\setminus T|}\le1
 &&(T\in\transversals(\mbb{H})),\\
&f^*(U)\ge0 &&(U\in\mintransset(\F)).
\end{aligned}
}\tag{DLP-1}\label{eq:dlp1}
\end{equation}
This is a packing LP: each set $U$ receives weight $f^*(U)$, earns value
$r^{|U|}$ per unit weight, and for every restriction $T\subseteq U$ uses $2^{-|U\setminus T|}$ units of its capacity.

The value of~\eqref{eq:dlp1} is exactly $Z(\F,r)$. Indeed, the assignment
$f^*(U)=1$ for all $U\in\mintransset(\F)$ is feasible by
Lemma~\ref{lem:restriction-weight} and has this value. Conversely,
$\mintransset(\F)\subseteq\transversals(\mbb{H})$, so the constraint indexed by $T=U$
implies $f^*(U)\le1$. No feasible assignment can have a larger objective.

The family $\mintransset(\F)$ still depends on $\G$. To obtain a bound determined
by $\mbb{H}$ alone, we retain the same restriction inequalities but allow a
variable for every $U\in\transversals(\mbb{H})$:
\begin{equation}
\boxed{
\begin{aligned}
\maxvalue{\mbb{H}}=\max\quad &\sum_{U\in\transversals(\mbb{H})}f^*(U)r^{|U|}\\
\text{subject to}\quad
&\sum_{\substack{U\in\transversals(\mbb{H})\\T\subseteq U}}
 f^*(U)2^{-|U\setminus T|}\le1
 &&(T\in\transversals(\mbb{H})),\\
&f^*(U)\ge0 &&(U\in\transversals(\mbb{H})).
\end{aligned}
}\tag{DLP-2}\label{eq:dlp2}
\end{equation}
Any feasible solution of~\eqref{eq:dlp1} extends to a feasible solution
of~\eqref{eq:dlp2} by assigning zero to the additional variables, with
no change in its objective. In particular, the indicator
$f^*(U)=\mathds{1}_{\{U\in\mintransset(\F)\}}$ is feasible for~\eqref{eq:dlp2}.
Consequently,
\[
\theta_t(\F)r^t\le Z(\F,r)\le \maxvalue{\mbb{H}}.
\]

To upper-bound the objective of~\eqref{eq:dlp2}, assign a nonnegative
multiplier $\lambda(T)$ to each restriction inequality. Adding the
weighted inequalities gives
\[
\sum_{U\in\transversals(\mbb{H})}f^*(U)
\left(\sum_{\substack{T\in\transversals(\mbb{H})\\T\subseteq U}}
\lambda(T)2^{-|U\setminus T|}\right)
\le\sum_{T\in\transversals(\mbb{H})}\lambda(T).
\]
The left-hand side dominates the packing objective whenever the
coefficient of each $f^*(U)$ is at least $r^{|U|}$. Minimizing the
resulting upper bound yields the dual program:
\begin{equation}
\boxed{
\begin{aligned}
\min\quad &\sum_{T\in\transversals(\mbb{H})}\lambda(T)\\
\text{subject to}\quad
&\sum_{\substack{T\in\transversals(\mbb{H})\\T\subseteq U}}
 \lambda(T)2^{-|U\setminus T|}\ge r^{|U|}
 &&(U\in\transversals(\mbb{H})),\\
&\lambda(T)\ge0 &&(T\in\transversals(\mbb{H})).
\end{aligned}
}\tag{C}\label{eq:raw-dual}
\end{equation}

We now express each multiplier in the same scale as the weighted sum:
write $\lambda(T)=f(T)r^{|T|}$ and put $q=(2r)^{-1}$. Dividing the
constraint for $U$ by $r^{|U|}$ gives
\[
\sum_{\substack{T\in\transversals(\mbb{H})\\T\subseteq U}}
 f(T)q^{|U|-|T|}\ge1,
\qquad
\frac{r^{|T|}2^{-|U\setminus T|}}{r^{|U|}}
 =\left(\frac1{2r}\right)^{|U|-|T|}.
\]
After renaming, we obtain
\begin{equation}
\boxed{
\begin{aligned}
\minvalue{\mbb{H}}=\min\quad &\sum_{U\in\transversals(\mbb{H})}f(U)r^{|U|}\\
\text{subject to}\quad
&\sum_{\substack{U\in\transversals(\mbb{H})\\U\subseteq S}}
 f(U)q^{|S|-|U|}\ge1
 &&(S\in\transversals(\mbb{H})),\\
&f(U)\ge0 &&(U\in\transversals(\mbb{H})).
\end{aligned}
}\tag{P}\label{eq:cover}
\end{equation}
This is a covering linear program which is dual to the linear program \ref{eq:dlp2}. Its coverage coefficient is forced by the restriction bound and the change of variables: $rq=1/2$.

The choice $f(U)=1$ for every $U\in\transversals(\mbb{H})$ is feasible for \eqref{eq:cover}, since the term $U=S$ already supplies one unit of coverage. Together with the feasibility and boundedness of \eqref{eq:dlp2}, finite-dimensional LP strong duality gives attained optima with $\maxvalue{\mbb{H}}=\minvalue{\mbb{H}}$. We have therefore proved
\[
\boxed{\theta_t(\F)r^t\le Z(\F,r)\le \maxvalue{\mbb{H}}=\minvalue{\mbb{H}}.}
\]
For the application, take $p=\sqrt{3/2}$, $r=p/2$, and $q=1/p$.
It remains to bound the covering cost
using only the structure of $\mbb{H}$: a bound $\minvalue{\mbb{H}}\le p^{n(\mbb{H})}$
would imply
$\theta_t(\F)\le r^{-t}p^{n(\mbb{H})}
=2^t p^{n(\F)-t}$,
since $n(\mbb{H})=n(\F)$.

\section{Main Theorem}

We will prepare the necessary ideas and notations for stating and proving the main theorem.
\subsection{Restrictions}

We consider restrictions where variables can be set to $0$, $1$, or left unset. In the following, we state our concepts and results generally in terms of set systems $\A$ over $V$. In our application of these concepts and results, we set $\A=\transversals(\mbb{H})$ of a monotone $3$-CNF.

\begin{definition}[Consistency of restrictions]
 Let $\rho: V \to \{0,1,*\}$ be a restriction. We say that  restrictions $\rho$  and $\rho'$ are consistent with each other if for all $x\in V$ whenever $\rho(x), \rho'(x) \in \{0,1\}$, $\rho(x)=\rho'(x)$. We say that a set $S\subseteq V$  is \emph{strongly consistent} with a restriction $\rho$ if its characteristic function $\mathds{1}_{S}$ is consistent with $\rho$.
\end{definition}

\begin{notation}
Let $\support(\rho) = \{v\mid \rho(v)\in \{0,1\}\}$,  
$\supportplus(\rho) = \{v\mid \rho(v)=1\}$, and $\supportnegative(\rho) = \{v\mid \rho(v)=0\}$.
\end{notation}

\begin{note}
Occasionally we will consider a slightly more general notion of restriction augmented by arbitrary clauses on the non-fixed variables. For example, for $V = \set{a, b, c, d}$, $\rho:= a = 1, b = 0, (c \vee d)$ is generalized restriction. To simplify notation we will write $cd$ instead of $c \vee d$. The notion of consistency of a set $S$ with restriction $\rho$ also naturally generalizes by additionally requiring that $S$ is a transversal of the augmenting clauses of $\rho$.
\end{note}

\begin{definition}
For a set $A\subseteq V$ strongly consistent with $\rho$, let $A_{\rho}$ denote the restriction of $A$ under $\rho$ defined as 
$A_{\rho} =A\setminus \supportplus(\rho)$. 
For a set system $\A$ on $V$ and a restriction $\rho$, define $\A_{\rho} = \{A_{\rho}\mid  A\in \A \text{ and $A$ is strongly consistent with $\rho$}\}$.
Define the  \emph{lift} $\lift_{\rho}(\B)$ of a set system $\B$ on $V\setminus \support(\rho)$ with respect to  $\rho$ as $\lift_{\rho}(\B) = \{B\sqcup \supportplus(\rho)\mid B\in \B\}$.
\end{definition}

\begin{fact}
$\lift_{\rho}(\A_{\rho})\subseteq \A$.
\end{fact}

\begin{definition}[Compatibility]
For a given set system $\A$, we say that a set $A\in \A$  is \emph{compatible} with a restriction $\rho$ if $\supportplus(\rho)\subseteq A$ and $A\setminus\support(\rho) \in \A_{\rho}$.
\end{definition}

\begin{definition}

We say that a set of restrictions  $\{\rho_i\}$ is complete if for every $A\in \A$ there exists an $i$ such that $A$ is compatible with $\rho_i$. 
\end{definition}

\begin{lemma}
Restrictions preserve minimal transversals:
    If $S\in \mintransset(\mbb{H})$ is strongly consistent with a restriction $\rho$, then $S\setminus \supportplus(\rho) \in \mintransset(\mbb{H}_{\rho})$.
\end{lemma}

\begin{definition}
    Let $\rho$ is a restriction and $\hat{f}$ is a weight function on $\A_{\rho}$, we define
    the lift $\lift_{\rho}(\hat{f})$ of $\hat{f}$  as follows. For $A\in \A$,
    \begin{align*}
        \lift_{\rho}(\hat{f})(A) &=
        \begin{cases}
            \hat{f}(A_{\rho}) & \text{if $A$ is strongly consistent with $\rho$ }\\
            0 & \text{if $A$ is not strongly consistent with $\rho$}
        \end{cases}
    \end{align*}
\end{definition}

It is easy to compute the cost of the lifted weights as follows.

\begin{fact}
$\cost(\lift_{\rho}(f_{\rho})) =  r^{\vert \supportplus(\rho) \vert }\cost(f_{\rho})$.
\end{fact}

In our applications we need to keep track of the coverage contributed by a restriction. This is given as follows.

\begin{fact}
Let $\rho$ be a restriction and $f_{\rho}$ be a feasible weight function on $\A_{\rho}$. For $A\in \A$ compatible with $\rho$, the contribution of $\lift_{\rho}(f_{\rho})$ towards the feasibility of $A$ is at least $q^{\vert A\cap \supportnegative(\rho)\vert}$.
\end{fact}

\begin{proof}
Write
\[
    P=\supportplus(\rho),\qquad
    N=\supportnegative(\rho),\qquad
    A_0=A\setminus\support(\rho).
\]
Since $A$ is compatible with $\rho$, we have $P\subseteq A$
and $A_0\in\A_\rho$. Moreover,
\[
    A=P\sqcup(A\cap N)\sqcup A_0.
\]
By the definition of the lift, its nonzero contributions to
the coverage of $A$ come from sets of the form $B=P\sqcup T$,
where $T\in\A_\rho$ and $T\subseteq A_0$. Hence
\begin{align*}
    \sum_{\substack{B\subseteq A\\B\in\A}}
    \lift_\rho(f_\rho)(B)q^{|A|-|B|}
    &=
    \sum_{\substack{T\subseteq A_0\\T\in\A_\rho}}
    f_\rho(T)q^{|A|-|P|-|T|}\\
    &=
    q^{|A\cap N|}
    \sum_{\substack{T\subseteq A_0\\T\in\A_\rho}}
    f_\rho(T)q^{|A_0|-|T|}\\
    &\ge q^{|A\cap N|}.
\end{align*}
The last inequality follows from the feasibility of $f_\rho$
for $\A_\rho$, applied to $A_0\in\A_\rho$.
\end{proof}


The following lemma is the main technical property needed to bound the cost function obtained from inductive bounds on cost functions of restrictions. In our application we will identify a set of restrictions along with scaling factors. Under the feasibility condition, we will lift the costs by their corresponding scaling factors and obtain a feasible weight and a cost bound for the original formula.

\begin{lemma}
Let $\A$ be a set system on a ground set $V$ of size $n$,
and let $\rho_1,\ldots,\rho_k$ be restrictions on $V$.
For each $A\in\A$, define $$I(A)=\{i\in\{1,\ldots,k\}: A \text{ is compatible with }\rho_i\}$$.
Let $\lambda_1,\ldots,\lambda_k\geq 0$ satisfy
\[
    \sum_{i\in I(A)}
    \lambda_i q^{|A\cap\supportnegative(\rho_i)|}
    \geq 1
    \qquad\text{for every }A\in\A.
\]
For each $i$, let $f_{\rho_i}$ be a feasible weight function
for $\A_{\rho_i}$. Then $f=\sum_{i=1}^k \lambda_i\lift_{\rho_i}(f_{\rho_i})$ is a feasible weight function for $\A$, and
$\cost(f) = \sum_{i=1}^k \lambda_i r^{|\supportplus(\rho_i)|} \cost(f_{\rho_i}).$ Furthermore, if for each $i$, $\primalopt{\A_{\rho_i}} \leq \omega_i p^{n-|\support(\rho_i)|}$, then
\begin{align*}
    \frac{\primalopt{\A}}{p^n}
    &\leq
    \sum_{i=1}^k
    \lambda_i\omega_i
    \frac{r^{|\supportplus(\rho_i)|}}
         {p^{|\support(\rho_i)|}}\\
    &=
    \sum_{i=1}^k
    \lambda_i\omega_i
    2^{-|\supportplus(\rho_i)|}
    q^{|\supportnegative(\rho_i)|}.
\end{align*}
\end{lemma}

\begin{proof}
Fix $A\in\A$. By nonnegativity and the preceding coverage fact,
\begin{align*}
    \sum_{\substack{B\subseteq A\\B\in\A}}
    f(B)q^{|A|-|B|}
    &=
    \sum_{i=1}^k \lambda_i
    \sum_{\substack{B\subseteq A\\B\in\A}}
    \lift_{\rho_i}(f_{\rho_i})(B)q^{|A|-|B|}\\
    &\geq
    \sum_{i\in I(A)}
    \lambda_i q^{|A\cap\supportnegative(\rho_i)|}\\
    &\geq 1.
\end{align*}
Thus $f$ is feasible for $\A$.

The cost identity follows from linearity of the objective
and the definition of the lift: each set receiving weight
from $f_{\rho_i}$ gains exactly
$|\supportplus(\rho_i)|$ elements when lifted.

For the final bound, choose each $f_{\rho_i}$ to attain
$\primalopt{\A_{\rho_i}}$. The resulting feasible function
has cost at most $\sum_{i=1}^k \lambda_i r^{|\supportplus(\rho_i)|} \omega_i p^{n-|\support(\rho_i)|}$. Divide by $p^n$ and use $r/p=1/2$, $1/p=q$, and $|\support(\rho_i)| = |\supportplus(\rho_i)|+|\supportnegative(\rho_i)|$ to obtain the claimed normalized bound.
\end{proof}









\subsection{Direct product of set systems}
For two set systems,  $\A_1$ and $\A_2$, we define their direct product 
$\A_1 \otimes \A_2$ as 
\begin{align*}
\A_1 \otimes \A_2 = \{A_1 \cup A_2 \mid A_1 \in \A_1, A_2\in \A_2\}.    
\end{align*}

\begin{fact}
    Let $\A_1$ and $\A_2$ be set systems over $V_1$ and $V_2$ respectively. Let $f_1$ and $f_2$ be feasible weight functions over $\A_1$ and $\A_2$ respectively. If $V_1$ and $V_2$ are disjoint, then the following hold.
    \begin{enumerate}
        \item $f_1\otimes f_2$ is feasible for $\A_1 \otimes \A_2$ where
    \begin{align*}
        f_1\otimes f_2(A) = f_1(A_1)f_2(A_2) \text{ where $A_1 = A\cap V_1$ and $A_2 = A\cap V_2$}
    \end{align*}
    \item If $f_1$ and $f_2$ are optimal weight functions for $\A_1$ and $\A_2$ respectively, then $f_1\otimes f_2$ is optimal for $\A_1 \otimes \A_2$.
    \item $\cost(f_1\otimes f_2) = \cost(f_1) \cost(f_2)$
    \end{enumerate} 
\end{fact}

\subsection{No variables and no clauses}

If a formula has no variables, then it must have an empty set of clauses. The empty set is its only transversal. Its weight is $1$.

\subsection {Isolated variables}

Let $\F$ be a CNF over $n$ variables and assume that the variable $x$ does not appear in $\F$.  We represent this \emph{isolated} variable $x$ by an empty set of clauses. Let $f_1$ be a feasible weight function for $\F$ restricted to all the variables except $x$, and $f_2$ be the weight function on $\{\emptyset, \{x\}\}$ defined by
\begin{align*}
    f_2(\emptyset) &= 1\\
    f_2(\{x\}) &= 1-q.
\end{align*}
Then $f_1\otimes f_2$ is feasible for $\F$. We have $\cost(f_1\otimes f_2) = \frac{p(1+q)}{2}\cost(f_1)$.

\subsection{Isolated 2-clause and 3-clause}
Let $\F = (ab)$. Then $f(\{a\}) = f(\{b\})=1$ and $f(\{a,b\})=0$ is an optimal weight function. Its cost is $2r$.

Let $\F = (abc)$. Then $f(\{a\}) = f(\{b\})=f(\{c\})=1$ and $f(\{a,b\})=f(\{b,c\})= f(\{a,c\})=f(\{a,b,c\})=0$ is an optimal weight function. Its cost is $3r$.

Let $\F = (ab)(bc)(ca)$. Then  $f(\{a,b\})=f(\{b,c\})= f(\{a,c\})=1$ and $f(\{a,b,c\})=0$ is an optimal weight function. Its cost is $3r^2$.

\begin{lemma}
If $a$ is a unit clause of $\mbb{F}$, then $\primalopt{\mbb{F}} = r\primalopt{\mbb{F}[a=1]}$. Consequently,
$\primalopt{\mbb{F}}/p^n = \frac{1}{2} \primalopt{\mbb{F}[a=1]}/p^{n-1}$.
\end{lemma}

Let $c_0 =1, c_1 = 1/(2-q), c_2 = c_1^2$, and $c_3 =r$. We have $1=c_0 > c_1 > c_2 > c_3 > \frac{1}{2}$.

\begin{theorem}[Simultaneous upper bounds]
    For every monotone $3$-CNF $\F$ on $n$ variables, we have
    \begin{enumerate}[label={S\arabic*}]
        \item $\minvalue{\F} \leq c_0 p^n$ \label[Theorem]{thm:S1}
         \item If $\F$ contains a $2$-clause, $\minvalue{\F} \leq c_1 p^n$ \label[Theorem]{thm:S2}
         \item If $\F$ contains two $2$-clauses, $\minvalue{\F} \leq c_2 p^n$ \label[Theorem]{thm:S3}
         \item If $\F$ contains three $2$-clauses, $\minvalue{\F} \leq c_3 p^n$ \label[Theorem]{thm:S4}
    \end{enumerate}
\end{theorem}

\subsection{Proof of the Main Theorem}

The proof goes through several cases depending on the structure of the formula. For each case we identify a set of restrictions that are complete with respect to the set of transversals, along with appropriate scaling factors. We then apply inductive bounds for the cost function of these restricted formulas and lift them back to the original formula using the tools discussed in the previous part.

\subsubsection{One 2-clause}
 Let $\F = (ab)\F'$. If neither $a$ nor $b$ appears in $\F'$, we have a direct product of two CNFs. There exists a feasible weight function for $(ab)$ such that the cost of $(ab)$ is $2r$. By induction, there exists a feasible weight function of $\F'$ with cost at most $p^{n-2}$ which implies that there is a feasible solution of $\F$ with cost at most $2rp^{n-2}$. The normalized cost is $q <c_1$ as desired.

 Otherwise, assume that $a$ appears in in the clause $(axy)$ where $x$ and $y$ are distinct from $b$. Each transversal of $\F$ is covered by one of two restrictions: $a=1$ or $a=0$ and $b=1$. For the first restriction, $\minvalue{\F[a=1]}=\minvalue{\F'[a=1]}\leq p^{n-1}$. For the second restriction, we have a $2$-clause in the restricted formula. By induction, $\minvalue{\F[a=0,b=1]}\leq c_1p^{n-2}$. After lifting the costs, the normalized $\minvalue{\F} \leq \frac{1}{2}+ c_1r/p^2 = \frac{1}{2}+ \frac{q}{2}c_1 = c_1$.
 
\subsubsection{Two 2-clauses}

Let $\F=(ab)(cd)\F'$. First assume that the $2$-clauses $(ab)$ and $(cd)$ overlap. Rename them as $(ab)$ and $(ac)$.   Consider the following two restrictions:
$a=1$ and $a=0, b=c=1$. For the first restriction, the lifted cost of $\F[a=1]$ is at most $p^{n-1}r$. For the second restriction, the value lifted cost is at most $p^{n-3}r^2$. The total normalized value is at most $\frac{1}{2}+\frac{q}{4}<c_2$.

Now assume that $(ab)$ and $(cd)$ are disjoint. If none of the $a,b,c$ and $d$ appear in $\F'$, we have two isolated $2$-clauses. The value of $\F$ is then bound by $4r^2p^{n-4}$. Its normalized value is bounded by $q^2<c_2$.

Otherwise, assume without loss of generality that $a$ appears in the clause $(axy)$ where $x$ and $y$ are distinct from $b$ and $(xy)$ is distinct from $(cd)$. Consider two restrictions $a=1$ or $a=0$ and $b=1$. The lifted cost of the first restriction is at most $c_1rp^{n-1}= c_1\frac{1}{2}p^n$. The lifted cost of the second restriction is at most
$c_2rp^{n-2} = c_2\frac{q}{2}p^n$. The total normalized cost is at most $\frac{1}{2}c_1 + \frac{q}{2}c_2 = c_2$.

\subsubsection{Three 2-clauses}

\paragraph{Star of three pairs.} Let $\F = (ab)(ac)(ad)\F'$. Consider the following two restrictions: $a = 1$ and $a = 0, b = c = d = 1$. Every transversal of $\F$ is consistent with exactly one of the two restrictions. By induction the normalized cost is at most $\frac{1}{2} + \frac{q}{8} < c_3$.

\paragraph{Triangle of three pairs.} Let $\F = (ab)(bc)(ac)\F'$. If one of the variables, say $a$, appears in $\F'$, we consider the following two restrictions $a = 1$ and $a = 0, b = c = 1$. The latter produces a pair. By induction, the normalized cost is at most $c_1(\frac{1}{2} + \frac{q}{4}) < c_3$.

We now assume that the triangle is isolated. We assign a weight of 1 to each of $ab$, $ac$, and $bc$. Observe that the isolated triangle has cost $3r^2$. By induction, the normalized cost of $\F$ is at most $\frac{3r^2}{p^3} = r = c_3$.

\paragraph{Path of three pairs.} Let $\F = (ab)(bc)(cd)\F'$. Assume $b$ appears in $bxy \in \F'$. Observe that $a, c \not \in \set{x,y}$. We consider two restrictions: $b = 1$ and $b = 0, a = c = 1$. Under $b = 1$, $cd$ survives and by induction we have $\minvalue{\F[b = 1]} = \minvalue{(cd)\F'[b = 1]} \leq  c_1 p^{n - 1}$. Under $b = 0, a = c = 1$ $xy$ survives. By induction we have $\minvalue{\F[b = 0, a = c = 1]} = \minvalue{(xy)\F'[b = 0, a = c = 1]} \leq c_1 p^{n - 3}$. After lifting the weights, the normalized cost is at most $c_1(\frac{1}{2} + \frac{q}{4}) < r$.

We now assume that neither $b$ nor $c$ appear in $\F'$. Consider the following restrictions.

\begin{align*}
    \rho_1 &:= b = d = 1, c = 0 \\
    \rho_2 &:= c = a = 1, b =0 \\
    \rho_3 &:= b = c = 1, a = d = 0 \\
    \rho_4 &:= a = b = c = 1, d = 0 \\
    \rho_5 &:= b = c = d = 1, a = 0.
\end{align*}

Consider a feasible weight $w_i$ that attains optimal cost for each $\rho_i$. We lift each weight $i$ with scale $\lambda_i$ defined as follows: $\lambda_1 = \lambda_2 = \lambda_3 = 1$ and $\lambda_4 = \lambda_5 = 1 - q$. We now show that the lifted weight $\sum_{i}\lift_{\rho_i}(w_i)$ is feasible. Let $A$ be any transversal. We consider all the possible ways in which $A$ may intersect $bc$. If $b \in A$ and $c \not \in A$, then $A$ is consistent with $\rho_1$, and since $\lambda_1 = 1$, $A$ is covered. If $b \not \in A$ and $c \in A$, coverage follows by a symmetric argument from $\rho_2$. 

We now assume that $b, c \in A$. If $a, d \not \in A$, $A$ is consistent with $\rho_3$, and since $\lambda_3 = 1$, $A$ is covered. We now assume that $a \in A$. Observe that $A \setminus \set{a, b, c}$ is a transversal of $\F[\rho_2]$. Since $|A \cap \supportnegative(\rho_2)| = 1$, $\rho_2$ contributes at least $\lambda_2 q = q$ to $A$'s coverage. If $d \not \in A$ then $A$ is consistent with $\rho_4$, and since $\lambda_4 = 1 - q$, $\rho_4$ contributes $1 - q$ to $A$'s coverage. In total $A$'s coverage is at least $q + 1 - q = 1$. By symmetry, this also shows that $A$ is covered when $d \in A$ and $a \not \in A$. Now consider the remaining case where $a, d \in A$. By a similar argument as above $\rho_1$ and $\rho_2$ each contribute $q$. Observe that $A \setminus \set{a, b, c, d}$ is a transversal of both $\F[\rho_4]$ and $\F[\rho_5]$. Therefore each contribute at least $(1 - q)q$. In total we get a contribution of $2q + 2(1 - q)q \ge 1$.

It now remains to upper bound the normalized cost. By induction, the lifted cost of each of $\rho_1$ and $\rho_2$ is $q/4$. The lifted cost of $\rho_3$ is $q^2/4$. The lifted cost of each of $\rho_4$ and $\rho_5$ is $(1 - q)q/8$. The total cost therefore is $2q/4 + q^2/4 + 2(1 - q)q/8 = q/2 + q/4 = r$.

\paragraph{Two adjacent pairs and a disjoint pair.} Let $\F = (ab)(bc)(de)\F'$. Consider restrictions $b = 1$, and $b = 0, a = c = 1$. In both we still retain $de$. The normalized cost is therefore $c_1(\frac{1}{2} + \frac{q}{4}) < c_3$.

\paragraph{Three disjoint pairs.} Let $\F = (ab)(cd)(ef)\F'$. If one of the variables, say $a$, appears in $\F'$ we consider the following restrictions $a = 1$, $a = 0, b = 1$. The former retains two pairs, and the latter retains three pairs. The normalized cost is therefore $c_2\frac{1}{2} + c_3\frac{q}{2} < c_3$. If the three pairs are isolated, the normalized cost is $p^3/p^6 = q^3 < c_3$.

\subsection{Proof of \ref{thm:S1}}

We now prove the statement \ref{thm:S1} of the main theorem for formulas over $n$ variables assuming all statements on smaller ground sets and statements \ref{thm:S2} through \ref{thm:S4}  on ground sets of size $n$.

An empty clause has $0$ value. A clause-free formula has cost at most $(r+\frac{1}{2})^n$.
Assume that every clause has exactly three variables.

We consider the following cases for $\F$ and consider the first case that $\F$ satisfies.

\begin{description}
\item[$\prop{0}{1}$:] \czeroI.
\item[$\prop{0}{2}$:] \czeroII.
\item[$\prop{0}{3}$:] \czeroIV.
\item[$\prop{0}{4}$:] \czeroV.
\item[$\prop{0}{5}$:] \czeroVI.
\end{description}

It is not hard to verify that for every $\mbb{F}$ with no clause of width less than three, one of the properties $\prop{0}{1}$ through $\prop{0}{5}$ holds.




\subsubsection{A repeated pair}
Let $\F$ contain the clauses $(abc)$ and $(abd)$. We consider two cases.

\paragraph{One of $a$ or $b$ appears in a clause without the other.}
Assume that $a$ appears in $(axy)$ where neither $x$ nor $y$ is equal to $b$. Consider the restrictions $a=1$ and $a=0$. The lifted cost of the first restriction is at most $rp^{n-1}$ and the lifted cost of the second restriction is at most $c_3 p^{n-1}$. The total normalized cost is at most $\frac{r}{p}+ c_3\frac{1}{p}=1$.

\paragraph{$a$ and $b$ always appears together.}
Let $F= (abc)(abd)\F'$ where $\F'$ contains neither $a$ nor $b$. We consider three restrictions:$a=1,b=0$, $a=0,b=1$, and $a=b=0$. Any transversal that contains both $a$ and $b$ gets a coverage of $q$ from each of the first two restrictions, and thus a total of $2q > 1$. For the first restriction, the lifted cost is $rp^{n-2}$. The second restriction is similar. For the third restriction, the lifted cost is at most $r^2p^{n-4}$. The total normalized cost is at most $\frac{2r}{p^2}+ \frac{r^2}{p^4} = q+\frac{1}{6}<1$.

\subsubsection{A variable of degree at least three}

Assume that $\F$ contains the clauses $(abc)$, $(ade)$ and ($afg)$.
Since $\F$ does not contain a repeated pair, $b,c,d,e,f$ and $g$ are distinct. We consider the restrictions $a=0$ and $a=1$.  The lifted cost for the first restriction is at most $c_3 p^{n-1}$ and the cost for the second restriction is at most $rp^{n-1}$. The total normalized cost is at most $c_3 q +\frac{1}{2} = 1$.

\subsubsection{Triangle configuration}

By the preceding cases, every variable occurs at most twice and no pair occurs in two distinct clauses. Write $\F=(abd)(bce)(acf)\F'$, where the six displayed variables are distinct. The variables $a,b,c$ do not occur in $\F'$, and each of $d,e,f$ occurs in $\F'$ at most once.

Set
\[
    \delta=1-q,\qquad \beta=\frac{1+q}{2}.
\]
Recall that an isolated variable has a feasible weight function
of cost $p\beta$.

\paragraph{Restrictions:}

We use the following restrictions and scaling factors:
\begin{align*}
    \rho_1
    &\coloneqq a=b=c=0,\ d=e=f=1,
    & \lambda_1&=1,\\
    \rho_2
    &\coloneqq a=1,\ b=c=0,\ e=1,
    & \lambda_2&=1,\\
    \rho_3
    &\coloneqq b=1,\ a=c=0,\ f=1,
    & \lambda_3&=1,\\
    \rho_4
    &\coloneqq c=1,\ a=b=0,\ d=1,
    & \lambda_4&=1,\\
    \rho_5
    &\coloneqq a=b=1,\ c=0,\ e=f=0,
    & \lambda_5&=1,\\
    \rho_6
    &\coloneqq a=b=1,\ c=0,\ e=1,\ f=0,
    & \lambda_6&=\delta,\\
    \rho_7
    &\coloneqq a=b=1,\ c=0,\ e=0,\ f=1,
    & \lambda_7&=\delta,\\
    \rho_8
    &\coloneqq b=c=1,\ a=0,\ d=f=0,
    & \lambda_8&=1,\\
    \rho_9
    &\coloneqq b=c=1,\ a=0,\ d=1,\ f=0,
    & \lambda_9&=\delta,\\
    \rho_{10}
    &\coloneqq b=c=1,\ a=0,\ d=0,\ f=1,
    & \lambda_{10}&=\delta,\\
    \rho_{11}
    &\coloneqq a=c=1,\ b=0,\ d=e=0,
    & \lambda_{11}&=1,\\
    \rho_{12}
    &\coloneqq a=c=1,\ b=0,\ d=1,\ e=0,
    & \lambda_{12}&=\delta,\\
    \rho_{13}
    &\coloneqq a=c=1,\ b=0,\ d=0,\ e=1,
    & \lambda_{13}&=\delta.
\end{align*}
All variables not specified by a restriction remain unset.


\paragraph{Coverage:}

Let $S$ be a transversal of $\F$.
We first record a sufficient condition for a restriction $\rho_i$
to contribute towards its coverage:
\[
    \supportplus(\rho_i)\subseteq S
    \qquad\text{and}\qquad
    S\cap\supportnegative(\rho_i)\subseteq\set{a,b,c}.
\]
Under these conditions,
\[
    S_i=S\setminus\support(\rho_i)
\]
is a transversal of $\F[\rho_i]$.
Indeed, the displayed triangle clauses are satisfied by the variables
set to one, and no variable of $S$ occurring in $\F'$ is set to zero.

For every $T\subseteq S_i$, we have
\[
    \card{S}-\card{T\cup\supportplus(\rho_i)}
    =
    \card{S_i}-\card{T}
    +\card{S\cap\supportnegative(\rho_i)}.
\]
Feasibility of $w_i$ therefore shows that the scaled lift contributes
at least
\[
    \lambda_iq^{\card{S\cap\supportnegative(\rho_i)}}.
\]

We now consider $P=S\cap\set{a,b,c}$.

\begin{itemize}
    \item
    If $P=\emptyset$, then $\set{d,e,f}\subseteq S$.
    The restriction $\rho_1$ contributes one.

    \item
    If $\card{P}=1$, assume without loss of generality that
    $P=\set{a}$.
    Since $S$ meets $bce$, we have $e\in S$.
    The restriction $\rho_2$ contributes one.
    The other two cases are symmetric.

    \item
    If $\card{P}=2$, assume without loss of generality that
    $P=\set{a,b}$.
    The following table gives sufficient contributions in all four
    possibilities for $S\cap\set{e,f}$:
    \[
    \begin{array}{c|c|c}
        S\cap\set{e,f}
        & \text{Contributing restrictions}
        & \text{Coverage}\\ \hline
        \emptyset & \rho_5 & 1\\
        \set{e} & \rho_2,\rho_6 & q+\delta=1\\
        \set{f} & \rho_3,\rho_7 & q+\delta=1\\
        \set{e,f} & \rho_2,\rho_3 & 2q\ge1
    \end{array}
    \]
    The other choices of $P$ are symmetric.

    \item
    Suppose that $P=\set{a,b,c}$.
    Up to cyclic relabelling, the possibilities are as follows:
    \[
    \begin{array}{c|c|l}
        S\cap\set{d,e,f}
        & \text{Contributing restrictions}
        & \text{Coverage}\\ \hline
        \emptyset
        & \rho_5,\rho_8,\rho_{11}
        & 3q\\
        \set{d}
        & \rho_4,\rho_5,\rho_9,\rho_{12}
        & q^2+q+2\delta q=3q-q^2\\
        \set{d,e}
        & \rho_2,\rho_4,\rho_6,\rho_9
        & 2q^2+2\delta q=2q\\
        \set{d,e,f}
        & \rho_2,\rho_3,\rho_4
        & 3q^2=2
    \end{array}
    \]
    All four quantities are at least one.
\end{itemize}

Thus every transversal receives coverage at least one, and $w$ is
feasible.

\paragraph{Cost:}

Call a variable in $\set{d,e,f}$ \emph{active} if it occurs in $\F'$,
and \emph{inactive} otherwise.
An inactive variable that remains unset is isolated in the residual
formula. Factoring it off gives a factor $\beta$ in the normalized
cost.

The normalized lifted cost of $\rho_1$ is at most
\[
    \frac{r^3}{p^6}=\frac{q^3}{8}.
\]
Each of $\rho_2,\rho_3,\rho_4$ has normalized lifted cost at most
\[
    \frac{r^2}{p^4}=\frac{q^2}{4}=\frac16,
\]
with an additional factor $\beta$ for each inactive outer variable
left unset.

We next bound the combined cost of $\rho_5,\rho_6,\rho_7$.
These restrictions select $a,b$, set $c=0$, leave $d$ unset,
and assign $(e,f)$ the values $(0,0),(1,0),(0,1)$, respectively.
For $i,j\in\set{0,1}$, write
\[
    b_{ij}
    =
    \frac{\minvalue{\F'[e=i,f=j]}}{p^{n-5}}.
\]
Their combined normalized lifted cost is at most
\[
    \frac{r^2}{p^5}
    \bigl(b_{00}+\delta r(b_{10}+b_{01})\bigr)
    =
    \frac{q^3}{4}
    \bigl(b_{00}+\delta r(b_{10}+b_{01})\bigr).
\]

We first bound the expression in parentheses without using a possible
isolated-variable saving from $d$.
Set
\[
    U_1=c_1+\delta r(1+c_1),
    \qquad
    U_2=c_2+2\delta rc_1.
\]

\begin{itemize}
    \item
    If neither $e$ nor $f$ is active, the unrestricted induction
    bound gives $b_{00},b_{10},b_{01}\le1$.
    Hence the expression in parentheses is at most
    \[
        1+2\delta r=p.
    \]

    \item
    Suppose that exactly one of $e,f$ is active, say $e$.
    Setting $e=0$ exposes a $2$-clause from its unique occurrence
    in $\F'$.
    Since $f$ is inactive, this clause survives both assignments
    to $f$.
    Thus
    \[
        b_{00},b_{01}\le c_1,\qquad b_{10}\le1,
    \]
    and the expression in parentheses is at most $U_1$.

    \item
    Suppose that both $e,f$ are active and occur in distinct
    clauses of $\F'$.
    Neither of these clauses contains the other variable, because
    each variable has only one occurrence in $\F'$.
    Setting $e=f=0$ therefore exposes two distinct $2$-clauses.
    They are distinct because otherwise their original clauses
    would share a pair.
    Either mixed assignment exposes one $2$-clause.
    Consequently,
    \[
        b_{00}\le c_2,\qquad b_{10},b_{01}\le c_1,
    \]
    giving the bound $U_2$.

    \item
    Suppose that $e,f$ occur in the same clause of $\F'$.
    Setting $e=f=0$ produces a unit clause, so $b_{00}\le1/2$.
    Either mixed assignment satisfies their common clause, and
    the unrestricted bounds give $b_{10},b_{01}\le1$.
    The expression in parentheses is therefore at most
    \[
        \frac12+2\delta r\le U_2.
    \]
    Indeed,
    \[
        U_2-\left(\frac12+2\delta r\right)
        =\frac{31-27q}{50}>0.
    \]
\end{itemize}

If $d$ is inactive, it is isolated in all three residual formulas.
Factoring it off before applying the induction hypothesis multiplies
each of the preceding bounds by $\beta$.
Thus the combined normalized cost of $\rho_5,\rho_6,\rho_7$ is at most
$q^3/4$ times the appropriate bound $p,U_1$, or $U_2$, with an
additional factor $\beta$ when $d$ is inactive.

The same argument applies to $\rho_8,\rho_9,\rho_{10}$, with $e$
left unset and the pair $(d,f)$ assigned, and to
$\rho_{11},\rho_{12},\rho_{13}$, with $f$ left unset and the pair
$(d,e)$ assigned.

Using the definitions of $c_1,c_2$ and the identities
$q^2=2/3$ and $r=3q/4$, we obtain
\[
    U_1=\frac{27q-1}{20},
    \qquad
    U_2=\frac{3+24q}{25}.
\]
In particular,
\[
    U_1-U_2=\frac{39q-17}{100}>0.
\]

It remains to sum the costs according to the number of active outer
variables.

\begin{itemize}
    \item
    Suppose that none of $d,e,f$ is active.
    Each of $\rho_2,\rho_3,\rho_4$ leaves two inactive variables
    unset, giving total normalized cost at most $\beta^2/2$.
    Each group of three two-core restrictions has bound
    $\beta p$, giving total normalized cost at most
    \[
        \frac{3q^3}{4}\beta p=\frac{\beta}{2}.
    \]
    Including $\rho_1$, the total is at most
    \[
        \frac{\beta^2}{2}+\frac{\beta}{2}+\frac{q^3}{8}
        =
        \frac{11}{24}+\frac{7q}{12}<1.
    \]

    \item
    Suppose that exactly one of $d,e,f$ is active.
    Among $\rho_2,\rho_3,\rho_4$, one leaves two inactive variables
    unset and the other two each leave one.
    Their total normalized cost is at most
    $(\beta^2+2\beta)/6$.

    Of the three groups of two-core restrictions, one assigns
    the two inactive variables and has bound $p$.
    Each of the other two groups assigns one active and one
    inactive variable, while leaving an inactive variable unset,
    and therefore has bound $\beta U_1$.
    The total normalized cost is at most
    \[
        \frac{\beta^2+2\beta}{6}
        +\frac{q^3}{4}(p+2\beta U_1)
        +\frac{q^3}{8}
        =
        \frac{197}{360}+\frac{19q}{40}<1.
    \]

    \item
    Suppose that at least two of $d,e,f$ are active.
    We may omit all isolated-variable savings.
    The total normalized cost of $\rho_2,\rho_3,\rho_4$
    is at most $1/2$.

    If exactly two outer variables are active, one group of
    two-core restrictions has bound $U_2$ and the other two
    have bound $U_1$.
    If all three are active, all three groups have bound $U_2$,
    which is at most $U_1$.
    In either case, the total normalized cost is at most
    \[
        \frac12+\frac{q^3}{4}(U_2+2U_1)+\frac{q^3}{8}
        =
        \frac{68}{75}+\frac{13q}{150}<1.
    \]
\end{itemize}

The three strict inequalities follow, for example, from $q<49/60$.
Hence $w$ is a feasible weight function of cost less than $p^n$,
which completes the triangle case.

\subsection{$\refE$ configuration}

Let $\F=(abc)(ade)(bfg)(chi)\F'$.
By the preceding cases, the nine displayed variables are distinct,
every variable occurs at most twice, and no pair occurs in two clauses.
In particular, $a,b,c$ do not occur in $\F'$.

Write
\[
    X=\set{a,b,c},\qquad
    P_a=\set{d,e},\quad P_b=\set{f,g},\quad P_c=\set{h,i},
\]
and let $\delta=1-q$.

\paragraph{Restrictions:}

For every nonempty $A\subseteq X$ and $K\subseteq A$, define
$\rho_{A,K}$ by setting
\[
    x=1 \quad(x\in A),\qquad
    x=0 \quad(x\in X\setminus A),\qquad
    y=0 \quad\left(y\in\bigcup_{x\in K}P_x\right).
\]
All other variables remain unset. We use the following scaling factors,
which depend only on $\card{A}$ and $\card{K}$:
\[
\begin{array}{c|c|c}
\card{A} & \card{K} & \lambda_{A,K}\\ \hline
1 & 0 & 1\\
2 & 1 & \delta\\
2 & 2 & 2q-1\\
3 & 1 & \delta^2\\
3 & 2 & \delta(2q-1)\\
3 & 3 & 3\delta
\end{array}
\]
All unspecified scaling factors are zero. This gives nineteen
restrictions with nonzero scaling factors, all of which are positive.

For each such restriction, choose a feasible weight function
$f_{A,K}$ for $\F[\rho_{A,K}]$ using the induction hypothesis.
For every transversal $T$ of $\F[\rho_{A,K}]$, place weight
$\lambda_{A,K}f_{A,K}(T)$ on $T\cup A$.
Sum these lifted weight functions over all the restrictions.
Every set receiving weight is a transversal of $\F$.

\paragraph{Coverage:}

Let $S$ be a transversal of $\F$, and define
\[
    A_0=S\cap X,\qquad
    M=\set{x\in A_0:S\cap P_x=\emptyset}.
\]
Since $S$ meets $abc$, we have $A_0\ne\emptyset$.
Furthermore, if $S$ misses $P_x$, it must contain $x$.
Thus $M$ records all the arm pairs missed by $S$.

We claim that $\rho_{A,K}$ contributes at least
\[
    \lambda_{A,K}q^{\card{A_0}-\card{A}}
\]
towards the coverage of $S$ whenever
\[
    M\subseteq A\subseteq A_0
    \qquad\text{and}\qquad
    K\subseteq M.
\]
Indeed, $K\subseteq M$ means that none of the boundary variables
set to zero belongs to $S$.
The residual formula contains the pair $P_x$ for each $x\notin A$.
Since $M\subseteq A$, all these pairs are met by $S$.
The projection
\[
    S'=S\setminus\support(\rho_{A,K})
\]
also satisfies the restricted clauses of $\F'$.
Consequently, $S'$ is a transversal of $\F[\rho_{A,K}]$.

For every $T\subseteq S'$ receiving weight from $f_{A,K}$,
\[
    \card{S}-\card{T\cup A}
    =\card{S'}-\card{T}+\card{A_0}-\card{A}.
\]
Feasibility of $f_{A,K}$ therefore gives the claimed contribution.

Summing these contributions gives the following lower bounds:
\[
\begin{array}{c|c|l}
\card{A_0} & \card{M} & \text{Coverage}\\ \hline
1 & 0\text{ or }1 & 1\\
2 & 0 & 2q\\
2 & 1 & q+\delta=1\\
2 & 2 & 2\delta+(2q-1)=1\\
3 & 0 & 3q^2=2\\
3 & 1 & q^2+2q\delta+\delta^2=1\\
3 & 2 &
q\bigl(2\delta+(2q-1)\bigr)
  +2\delta^2+\delta(2q-1)=1\\
3 & 3 &
3\delta^2+3\delta(2q-1)+3\delta
  =3(1-q^2)=1
\end{array}
\]
Every entry is at least one, so the resulting weight function is feasible.

\paragraph{Cost:}

For $x\in X$, let $\alpha_x$ be the number of variables in $P_x$
that occur in $\F'$.
Thus $\alpha_x\in\set{0,1,2}$.
Every such variable occurs in exactly one clause of $\F'$.

Set
\[
    \beta=\frac{1+q}{2},
\]
so that an isolated variable contributes a factor $p\beta$.
For convenience, put $c_j=c_3$ for every $j\ge3$.

Fix a restriction $\rho_{A,K}$, and write
\[
    s(A,K)=3-\card{A}+\sum_{x\in K}\alpha_x,
    \qquad
    z(A,K)=\sum_{x\in A\setminus K}(2-\alpha_x).
\]
We claim that
\[
    \minvalue{\F[\rho_{A,K}]}
    \le
    \beta^{z(A,K)}c_{s(A,K)}
    p^{n-3-2\card{K}}.
\]

First, the residual formula contains the $3-\card{A}$ arm pairs
$P_x$ with $x\notin A$.
The restriction also sets
$\sum_{x\in K}\alpha_x$ variables occurring in $\F'$ to zero.

Suppose that no clause of $\F'$ contains two of these zeroed variables.
Their occurrences then lie in distinct clauses, each of which becomes
a $2$-clause.
These new $2$-clauses are distinct from one another and from the retained
arm pairs: an equality would give a pair occurring in two clauses of
$\F$.
The residual formula therefore contains $s(A,K)$ distinct $2$-clauses.

If instead a clause of $\F'$ contains at least two zeroed variables,
it becomes a unit clause or an empty clause.
An empty clause gives cost zero.
A unit clause gives normalized cost at most $1/2$, which is at most
$c_{s(A,K)}$.
Thus $c_{s(A,K)}$ is a valid coefficient in both cases.

Finally, each of the $z(A,K)$ variables counted above is isolated
in the residual formula: its arm is satisfied by a variable in $A$,
and it does not occur in $\F'$.
Factoring off these isolated variables supplies the additional factor
$\beta^{z(A,K)}$.
This also applies in the unit-clause case, since none of these variables
occurs in a clause of $\F'$.
The claimed bound follows by induction.

The restriction fixes $\card{A}$ variables to one and
$3-\card{A}+2\card{K}$ variables to zero.
Its scaled lift therefore has normalized cost at most
\[
    \frac{\lambda_{A,K}}{2^{\card{A}}}
    q^{3-\card{A}+2\card{K}}
    \beta^{z(A,K)}c_{s(A,K)}.
\]
Consequently, the total normalized cost is at most
\[
    C(\alpha_a,\alpha_b,\alpha_c)
    =
    \sum_{\substack{\emptyset\ne A\subseteq X\\K\subseteq A}}
    \frac{\lambda_{A,K}}{2^{\card{A}}}
    q^{3-\card{A}+2\card{K}}
    \beta^{z(A,K)}c_{s(A,K)}.
\]

This expression is symmetric in $\alpha_a,\alpha_b,\alpha_c$.
Using
\[
    q^2=\frac23,\qquad
    c_1=\frac{1}{2-q},\qquad
    c_2=c_1^2,\qquad
    c_3=r=\frac{3q}{4},
\]
we obtain the following ten cases:
\[
\begin{array}{c|c}
(\alpha_a,\alpha_b,\alpha_c)
    & C(\alpha_a,\alpha_b,\alpha_c)\\ \hline
(0,0,0) & (28669+11502q)/43200\\
(0,0,1) & (15226+12383q)/28800\\
(0,0,2) & (4473+25784q)/28800\\
(0,1,1) & (2844+4277q)/7200\\
(0,1,2) & (1601+7483q)/8640\\
(0,2,2) & (332+553q)/864\\
(1,1,1) & (1106+6423q)/7200\\
(1,1,2) & (157+762q)/864\\
(1,2,2) & (6262+4521q)/10800\\
(2,2,2) & (361-27q)/360
\end{array}
\]
Each entry is less than $19/20$.
For the first nine entries this follows by substituting $q<49/60$;
for the last entry it follows from $q>4/5$.
Hence the feasible weight function constructed above has cost less than
\[
    \frac{19}{20}p^n<p^n,
\]
which completes this case.

\subsection{Variable of degree one}

Let $\F = (abc)\F'$ and assume that $a$ occurs only in a triple $abc$.


\subsubsection{At least one of $b$ or $c$ also appears uniquely}
Suppose that $b$ appears uniquely. If $c$ also appears uniquely, then $abc$ is isolated. By induction, the total cost is $p^3*p^{n - 3} = p^n$.

We now assume that $c$ occurs in another clause $cde$. Since $a$ and $b$ appear uniquely, $d$ and $e$ are distinct from $a$ and $b$. We consider two restrictions: $c = 1$ and $c = 0$. Under $c = 1$, $a$ and $b$ are isolated variables. The total cost of this branch is $r\frac{p^2(1+q)^2}{4}p^{n - 3}$. The normalized cost is therefore $\frac{(1 + q)^2}{8}$.

Under $c = 0$, the resulting formula contains $ab$ and $de$ where $ab$ is isolated. The total cost is $pc_1p^{n - 3}$. The normalized cost is $c_1/p^2$.

The total normalized cost is $\frac{(1 + q)^2}{8} + \frac{c_1}{p^2} < 1$. 

\subsection{Both $b$ and $c$ have a second appearance}

Let $\F = (abc)(bde)(cfg)\F''$. Observe that $a$, $b$, and $c$ appear only in the clauses displayed. Furthermore, the pairs $de$ and $fg$ are disjoint and avoid $a,b,c$, otherwise we have a repeated pair or a triangle configuration. We define a complete set of restrictions along with their scaling factors as follows.

\begin{align*}
    \rho_1 &\coloneqq a = 1,\ b = c = 0;\ & \lambda_1&=1\\
    \rho_2 &\coloneqq b = 1,\ a = c = 0;\ & \lambda_2&=1\\
    \rho_3 &\coloneqq c = 1,\ a = b = 0;\ & \lambda_3&=1\\
    \rho_4 &\coloneqq a = b = 1,\ c = d = e = 0;\ & \lambda_4&=1 - q\\
    \rho_5 &\coloneqq a = c = 1,\ b = f = g = 0;\ & \lambda_5&=1 - q\\
    \rho_6 &\coloneqq b = c = 1,\ a = d = e = f = g = 0;\ & \lambda_6&=1\\
    \rho_7 &\coloneqq b = c = 1,\ a = d = e = 0; fg;\ & \lambda_7&=1 - q\\
    \rho_8 &\coloneqq b = c = 1,\ a = f = g = 0; de;\ & \lambda_8&=1 - q\\
    \rho_9 &\coloneqq a = b = c = 1,\ d = e = f = g = 0;\ & \lambda_9&=1 - q\\
    \rho_{10} &\coloneqq a = b = c = 1,\ d = e = 0; fg;\ & \lambda_{10}&=(1 - q)^2\\
    \rho_{11} &\coloneqq a = b = c = 1,\ f = g = 0; ed;\ & \lambda_{11}&=(1 - q)^2
\end{align*}

We first show that these restrictions cover all transversals. Let $A$ be a transversal and let $P = A \cap \set{a, b, c}$. Write $I = 1[A \cap \set{d, e} \ne \emptyset]$ and $J = 1[A \cap \set{f, g} \ne \emptyset]$. We consider several cases.

\begin{itemize}
    \item $|P| = 1$: $A$ is consistent with one of $\rho_1$, $\rho_2$, or $\rho_3$ nad gets full coverage.
    
    \item $P = \set{a, b}$ or $P = \set{a, c}$: Wlog, assume the former. If $I = 1$, then we claim that $A$ receives a coverage of $q$ from each of $\rho_1$ and $\rho_2$, and since $2q \ge 1$, it receives full coverage. To see this, observe that $A \setminus \set{a, b}$ is a transversal of both $\F[\rho_1]$ and $\F[\rho_2]$. Now assume $I = 0$. since $A \setminus \set{a, b}$ is a transversal of $\F[\rho_2]$, $\rho_2$ contributes $q$ to coverage. Furthermore, $A$ is consistent with $\rho_4$ which constributes $1 - q$. Overall we have a coverage of $q + 1 - q = 1$. The case of $P = \set{a, c}$ is symmetric.

    \item $P = \set{b, c}$. If $I = J = 1$, then each of $\rho_2$ and $\rho_3$ constributes $q$, since $A \setminus \set{b, c}$ is a transversal of $\F[\rho_2]$ and $\F[\rho_3]$. We have a total of $2q \ge 1$ coverage. Now assume $I = 0, J = 1$ (respectively $I = 1, J = 0$). $A$ is consistent with $\rho_7$ (respectively $\rho_8$) and receives a coverage of $1 - q$. $A$ also receives a coverage of $q$ from $\rho_2$ (respectively $\rho_3$), and thus an overal coverage of $q + 1 - q = 1$. Finally assume $I = J = 0$. In this case $A$ is consistent with $\rho_6$ from which it gets full coverage.

    \item $P = \set{a, b, c}$. If $I = J = 1$, then each of $\rho_1, \rho_2, \rho_3$ contributes $q^2$, and thus a total of $3q^2 = 2$. Now assume $I = 0, J = 1$ (respectively $I = 1, J = 0$). Then $\rho_2$ (respectively $\rho_3$) contributes $q^2$. Furthermore, $\rho_4$ and $\rho_7$ (respectively $\rho_5$ and $\rho_8$) together contribute $2q(1 - q)$. $A$ is consistent with $\rho_{10}$ (respectively $\rho_{11}$) from which it gets a coverage of $(1 - q)^2$. The total coverage is thus $q^2 + 2q(1 - q) + (1 - q)^2 = 1$. Finally, if $I = J = 0$, then $A$ is consistent with $\rho_9$ from which it gets a coverage of $1 - q$ and a coverage of $q$ from $\rho_6$, which is a total of coverage 1.
\end{itemize}

It remains to bound the cost. It is easy to verify that the normalized lifted cost for each of the restrictions is given as follows $c(\rho_1) = rc_2q^3$, $c(\rho_2) = c(\rho_3) = rc_1q^3$, $c(\rho_4) = c(\rho_5) = c(\rho_7) = c(\rho_8) = c_1(1 - q)r^2q^5$, $c(\rho_6) = r^2q^7$, $c(\rho_9) = (1 - q)r^3q^7$, and $c(\rho_{10}) = c(\rho_{11}) = c_1(1 - q)^2r^3q^5$. Simple calculation shows that the total sum is less than 1.

\section{Algorithmic aspect}
\label{sec:algorithmic_aspect}

We now observe that our proof can be interpreted as a polynomial time sampling algorithm in the following sense: given a formula $\F$ with threshold $t$, in polynomial time the algorithm outputs each $t$-transversal with probability at least $r^t/p^n$. For this, we first design an algorithm, called $\Sample$, to sample the transversals of a monotone $3$-CNF using the proof of the main theorem constructively. We then apply $\Sample$ to the monotone part $\mbb{H}$ of $\mbb{F}$. If $\Sample$ outputs a transversal $T$, we extend it to a transversal of size $t$ if $\card{T}\leq t$. We then argue that each $t$-transversal of $\mbb{F}$ is output with a probability of at least $r^t/p^n$.

Recall that the proof takes a monotone formula $\mbb{H}$ and depending on the presence of various constant-size local configurations, it identifies a small set of restrictions and scaling factors. We note that this process can be performed in polynomial time. We process the formula in a particular order: if it has at least three 2-clauses we invoke the proof of \ref{thm:S4}, if it has exactly two 2-clauses we invoke \ref{thm:S3}, if it has exactly one 2-clause we invoke \ref{thm:S2}. Finally, if it has no 2-clause we invoke \ref{thm:S1}. For each of these, we process the formula in the order presented by the proof. For example, when invoking \ref{thm:S1}, we first check if there is a pair of variables appearing in at least two 3-clauses. If there is no such pair, then we check if there is a variable that appears at least three times, etc.

Let $\mbb{H}$ be the monotone part of $\mbb{F}$. The proof of the main theorem constructively defines a weight function $f$ such that $\sum_{T \in \transversals(\mbb{H})}f(T)r^{|T|} \le p^n$. We first show how to use the proof to sample transversals of $\mbb{H}$.

Given a monotone $3$-CNF $\mbb{H}$ on $n$ variables, the inductive proof supplies a polynomial-time computable upper bound $B(\mbb{H}) \le p^n$ on the cost of the feasible weight function constructed for $\mbb{H}$. We call $B(\mbb{H})$ its \emph{budget}. These budgets account for the elementary reductions used in the proof, including unit clauses, isolated variables, and disjoint components. In particular, at each branching step they satisfy
\[
    \sum_i \lambda_i r^{|\supportplus(\rho_i)|}
    B(\mbb{H}[\rho_i])
    \le B(\mbb{H}),
\]
where the restrictions $\rho_i$ and scaling factors $\lambda_i$ are those supplied by the corresponding case of the proof. This inequality ensures that the branch-selection probabilities defined below sum to at most one. We will give the full case-by-case definition of $B(\mbb{H})$ in the full version. For illustration, when none of the elementary reductions applies, the budget is given by the following expression.

$$
B(\mbb{H}) = 
\begin{cases}
p^n & \mbb{H} \text{ has no 2-clauses} \\
c_1 p^n & \mbb{H} \text{ has exactly one 2-clause} \\
c_2 p^n & \mbb{H} \text{ has exactly two 2-clauses} \\
c_3 p^n & \mbb{H} \text{ has at least three 2-clauses.}
\end{cases}
$$

We can now describe the sampling algorithm $\Sample(\mbb{H})$ where $\mbb{H}$ is a monotone $3$-CNF in $n$ variables
with budget $B(\mbb{H})$.

\begin{enumerate}
    \item If $\mbb{H}$ is empty then output $\emptyset$.
    \item If $\mbb{H}$ has an empty clause then return failure.
    \item Otherwise, let $\rho_1, \ldots, \rho_k$ be the set of restrictions given by the proof with scaling factor $\lambda_i$, and let $B_i = B({\mbb{H}[\rho_i]})$ for each $i \in [k]$.
    \item Choose $\rho_i$ with probability $\lambda_i r^{|\supportplus(\rho_i)|}B_i/B(\mbb{H})$, or return failure with the remaining probability.
    \item Recursively sample from $\mbb{H}[\rho_i]$ by running $\Sample(\mbb{H}[\rho_i])$ with budget $B(\mbb{H}[\rho_i])$. If this fails, return failure, otherwise return $\supportplus(\rho_i) \cup \Sample(\mbb{H}[\rho_i])$.
\end{enumerate}

It is immediate that when the algorithm outputs a set $S$ then it is indeed a transversal. We now describe the main algorithm. Given $\F$ with threshold $t$ and monotone part $\mbb{H}$, we run $\Sample(\mbb{H})$. If it fails we output failure. Otherwise let $A$ be the output. If $|A| > t$ output failure. While $|A| < t$, pick a clause in $\mbb{F}$ that is falsified by $A$. If it has no positive literal then reject, otherwise pick one of its positive literals uniformly at random and add to $A$ and repeat. Finally when $|A| = t$ output $A$ if it satisfies $\F$, otherwise output failure.

We now show a lower bound on the probability of outputting any given transversal $S$ (not necessarily a $t$-transversal) of $\mbb{F}$. Observe that this probability is at least $$\delta_{\mbb{H}}(S) := \sum_{T \in \transversals(\mbb{H}): T \subseteq S}\Pr[\Sample(\mbb{H}) = T]2^{-|S \setminus T|}.$$ 

For every $T \in \transversals(\mbb{H})$ such that $T \subseteq S$, since $T$ satisfies every clause in $\mbb{H}$, every remaining clause in $\F$ restricted to $T$ has at most 2 positive variables. Therefore, conditioned on that the output of $\Sample(\mbb{H})= T$, the probability that the algorithm produces $S$ is at least $2^{-|S\setminus T|}$.

It is therefore sufficient to prove the following lemma.

\begin{lemma}
Let $S \in \transversals(\mbb{H})$. Then $\sum_{T \in \transversals(\mbb{H}): T \subseteq S}\Pr[\Sample(\mbb{H}) = T]2^{-|S \setminus T|} \ge r^{|S|}/B(\mbb{H})$.
\end{lemma}

\begin{proof}
We prove the lemma by induction on the number of variables in $\mbb{H}$. Write $B = B(\mbb{H})$. Define $$I(S) = \{i: \supportplus(\rho_i) \subseteq S, S \setminus \support(\rho_i) \in \transversals(\mbb{H}[\rho_i])\}.$$ We have 
\begin{eqnarray*}
\delta_{\mbb{H}}(S) &=& \sum_{T \in \transversals(\mbb{H}): T \subseteq S}\Pr[\Sample(\mbb{H}) = T]2^{-|S \setminus T|}\\
&=& \sum_{T \in \transversals(\mbb{H}): T \subseteq S}2^{-|S \setminus T|}\sum_{i: \supportplus(\rho_i) \subseteq T, \supportnegative(\rho_i) \cap T = \emptyset} \Pr[\rho_i]\Pr[\Sample(\mbb{H}[\rho_i]) = T \setminus \supportplus(\rho_i)]\\
&=& \sum_{i \in I(S)}\Pr[\rho_i]\sum_{T \subseteq S: \supportplus(\rho_i) \subseteq T, \supportnegative(\rho_i) \cap T = \emptyset}\Pr[\Sample(\mbb{H}[\rho_i]) = T \setminus \supportplus(\rho_i)]2^{-|S \setminus T|}\\
&=& \sum_{i \in I(S)} \Pr[\rho_i]\sum_{T' \in \transversals(\mbb{H}[\rho_i]): T' \subseteq S \setminus \support(\rho_i)}\Pr[\Sample(\mbb{H}[\rho_i]) = T']2^{-|(S \setminus \support(\rho_i)) \setminus T'|} 2^{-|S \cap \supportnegative(\rho_i)|}\\
&=& \sum_{i \in I(S)} \frac{\lambda_i r^{|\supportplus(\rho_i)|}B_i}{B}2^{-|S \cap \supportnegative(\rho_i)|}\delta_{\mbb{H}[\rho_i]}(S\setminus \support(\rho_i))\\
&\ge& \sum_{i \in I(S)} \frac{\lambda_i r^{|\supportplus(\rho_i)|}B_i}{B}2^{-|S \cap \supportnegative(\rho_i)|}\frac{r^{|S \setminus \support(\rho_i)|}}{B_i}\\
&=& \frac{r^{|S|}}{B}\sum_{i \in I(S)} \lambda_i q^{|S \cap \supportnegative(\rho_i)|}\\
&\ge& \frac{r^{|S|}}{B},
\end{eqnarray*}
where the latter inequality follows by coverage.
\end{proof}

For $t$-transversal $S$ of $\F$, we have $\delta_{\mbb{H}}(S) \ge r^t/p^n$ as desired.

\section{Conclusions} 

The first obvious problem left open by our results is to extend the current techniques to $\Sigma_3^k$ for fixed $k\ge4$. The unrestricted asymptotic depth-three complexity of Majority is settled by \cite{LecomteRamakrishnan2026}; the remaining circuit questions concern restricted models and precise bounds, not the formerly conjectured super-$2^{\Theta(\sqrt n)}$ lower bound. There are several obstacles. First, it is not clear whether extremal constructions for $k \ge 4$ follow the modularity principle. Even if they do, it is not clear how to uniformly find the building blocks. We suspect \turanl-type building blocks will play a central role in these extremal constructions.

It would also be interesting to extend even the extremal construction for thresholds beyond $n/2$ even for monotone $\Sigma_3^3$. While we have certain guesses for the extremal constructions, proving their optimality for all thresholds, especially at threshold $t = n - 3$, will provably involve the resolution of famous hypergraph Tur{\'a}n problems.

We hope our techniques for characterizing extremal $3$-CNFs with respect to Boolean threshold functions (where the threshold $\leq \frac{n}{2}$) and their connections to extremal combinatorics inspire further work towards resolving various natural open problems. While precise characterizations of $k$-CNFs are rare, we do believe that inductive arguments of the style employed in this paper are powerful and can lead to optimal bounds  for $k>3$.
 

It is also interesting to find a shorter version (or a simpler alternative) of our proof. It is conceivable an intelligent computer search might be able to discover a shorter version of the current proof.

\section{AI disclosure}
In an earlier version of the paper we only had the result for monotone formulas. We used GPT 5.6 and 6 Astra to explore ideas which eventually led to the extension to non-monotone formulas. While the inductive proof closely mimics our original monotone proof, the LLM crucially found the right parameterization to make the argument go through.




\printbibliography

\end{document}